\documentclass[showpacs,preprint,preprintnumbers
,nofootinbib,prd,
eqsecnum,10pt
]{revtex4-1}

\usepackage{amsmath,amssymb}
\usepackage[normalem]{ulem}
\usepackage{textcomp}
\usepackage{hyperref}
\usepackage{bm}
\usepackage[bottom]{footmisc}
\usepackage{tensor}
\usepackage{graphicx}
\usepackage{psfrag}
\usepackage[usenames,dvipsnames]{xcolor}
\usepackage[utf8]{inputenc}
\usepackage[english, capitalise]{cleveref}
\crefname{chapter}{Chap.}{Chap.}
\crefname{section}{Sec.}{Sec.}
\Crefname{chapter}{Chapter}{Chapters}
\Crefname{section}{Section}{Sections}
\Crefname{eqs}{Eqs.}{Eqs.}
\Crefformat{eqs}{Eqs.~(#2#1#3)}

\definecolor{darkgreen}{rgb}{0,0.5,0}

\hypersetup{
    bookmarks=true,         
    unicode=false,          
    pdftoolbar=true,        
    pdfmenubar=true,        
    pdffitwindow=false,     
    pdfstartview={FitH},    
    pdftitle={My title},    
    pdfauthor={Author},     
    pdfsubject={Subject},   
    pdfcreator={Creator},   
    pdfproducer={Producer}, 
    pdfkeywords={keyword1} {key2} {key3}, 
    pdfnewwindow=true,      
    colorlinks=true,       
    linkcolor=red,          
    citecolor=cyan,        
    filecolor=magenta,      
    urlcolor=darkgreen,           
    linktocpage=true
}

\newcommand{\dd}{\mathrm{d}}
\newcommand{\dD}{\mathrm{D}}
\newcommand{\di}{\mathrm{i}}
\newcommand{\de}{\mathrm{e}}
\newcommand{\m}{\mathfrak{m}}

\newcommand{\hatG}{\hat{G}}
\newcommand{\hatH}{\hat{H}}

\allowdisplaybreaks

\begin{document}

\title{Tidal effects in the gravitational-wave phase evolution of compact binary systems to next-to-next-to-leading post-Newtonian order}

\author{Quentin \textsc{Henry}}
 \email{henry@iap.fr}
 \author{Guillaume \textsc{Faye}}
 \email{faye@iap.fr}
 \author{Luc \textsc{Blanchet}}
 \email{luc.blanchet@iap.fr}
\affiliation{$\mathcal{G}\mathbb{R}\varepsilon{\mathbb{C}}\mathcal{O}$, Institut d'Astrophysique de Paris, UMR 7095 CNRS,\\ Sorbonne Université, 98bis boulevard Arago, 75014 Paris, France}

\date{\today}

\begin{abstract}
We compute the gravitational-wave (GW) energy flux up to the next-to-next-to-leading (NNL) order of tidal effects in a spinless compact binary system on quasi-circular orbits. Starting from an effective matter action, we obtain the stress-energy tensor of the system, which we use in a GW generation formalism based on multipolar-post-Minkowskian (MPM) and post-Newtonian (PN) approximations. The tidal contributions to the multipole moments of the system are first obtained, from which we deduce the instantaneous GW energy flux to NNL order (formally 7PN order).  We also include the remaining tidal contributions of GW tails to the leading (formally 6.5PN) and NL (7.5PN) orders. Combining it with our previous work on the conservative equations of motion (EoM) and associated energy, we get the GW phase and frequency evolution through the flux-balance equation to the same NNL order. These results extend and complete several preceding results in the literature.
\end{abstract}

\maketitle

\section{Physical discussion and motivations}
\label{sec:intro}

The discovery of gravitational waves (GW) generated by the inspiral and merger of two neutron stars (NS)~\cite{GW170817, scientific2001gw190425} marked a breakthrough in fundamental physics, by allowing for the first time a direct constraint on the equation of state (EoS) of cold matter at supranuclear densities deep inside NS. This important test excluded some of the stiffest EoS, for which the pressure increases a lot for a given increase in density, and which therefore offer more resistance to the gravitational collapse, resulting in a NS that is less compact. This finding is consistent with known constraints on the radius of NS from electromagnetic-based observations~\cite{LP16}. However the majority of soft EoS, which are more easy to compress and thus predict a more compact NS, is still viable (see~\cite{BuonSathya15, DHS20} for reviews).

During the inspiral phase of coalescing NS binaries the orbital dynamics is dominated by point-mass contributions and the waveform is essentially identical to that of black holes; but closer to the merger small corrections arise due to the finite-size effects of NS. These can be described by resorting to a tidal expansion in the small parameter $\sim R_A/r_{AB}$, where $R_A$ is the size of one of the NS and $r_{AB}$ is the typical orbital separation. The tides arise from the response of the NS to the gradient of the companion's gravitational field across the matter distribution. The tidal expansion is a multipole expansion with the mass quadrupole moment of the object dominant, and higher mass- or current-type moments sub-dominant. The deformation and finite size effects are parametrized by a series of coefficients associated with each multipole moments and referred to as the tidal deformabilities (or polarizabilities) of the NS. 

For GW detectors, the main observable is the so-called binary's chirp, \textit{i.e.}, the time evolution of the compact binary's orbital frequency $\omega(t)$ and phase $\varphi(t)=\int\dd t\,\omega(t)$ through GW radiation reaction during the inspiral. The detectors are sensitive to some particular combination of the two deformabilities of the NS and the two masses that enters the binary's chirp. To the lowest tidal mass quadrupole order, the chirp is given by the combination of the two relations,
\begin{subequations}\label{eq:chirp}
	\begin{align}
	x &= \frac{1}{4}\theta^{-1/4}\left[ 1 + \frac{39}{8192} \tilde{\Lambda}^{(2)} \,\theta^{-5/4}\right]\,,\\
	\varphi &= \varphi_0 - \frac{x^{-5/2}}{32\nu} \left[ 1 + \frac{39}{8} \tilde{\Lambda}^{(2)} \,x^{5}\right]\,,
	\end{align}
\end{subequations}
where $\nu=m_1 m_2/m^2$ is the symmetric mass ratio, $m=m_1+m_2$ is the total mass, and where we use the dimensionless frequency $x \equiv (\frac{G m \omega}{c^3})^{2/3}$ and time $\theta \equiv \frac{\nu c^3}{5G m}(t_\text{c}-t)$ variables, with $t_\text{c}$ denoting the instant of coalescence --- at which the distance between the particles formally vanishes while the frequency diverges ---, $\varphi_0$ being an initial constant phase, $G$ the gravitational constant, and $c$ the speed of light. The most commonly used approximant for GW data analysis in the Fourier domain is defined by using the stationary phase approximation (SPA), for which the phase of the dominant mode at twice the orbital frequency reads
\begin{equation}\label{eq:phaseSPA}
\psi = 2 \pi f \,t_\text{c} -2 \varphi_0 - \frac{\pi}{4} + \frac{3 v^{-5}}{128\nu} \left[ 1 - \frac{39}{2} \tilde{\Lambda}^{(2)} \,v^{10}\right]\,,
\end{equation}
where we have posed $v \equiv (\frac{\pi G m f}{c^3})^{1/3}$, $f$ being the Fourier frequency of the GW signal. 

Since $x$ and $v^2$ are small post-Newtonian (PN) parameters of the order of $\mathcal{O}(c^{-2})$, we see that the effect of the internal structure of NS (in the non-spinning case) is comparable to relativistic orbital effects occurring at the 5PN order beyond the point-particle contribution computed with the usual Einstein quadrupole formula. Of course, the latter estimate is just formal, since we have to take into account, besides the small factor $x^5\sim v^{10}$, the numerical value of the 5PN coefficient parametrizing the finite-size effect in~\eqref{eq:chirp}--\eqref{eq:phaseSPA}.

In principle the coefficient $\tilde{\Lambda}^{(2)}$ is directly measurable by the GW detectors. However, in practice the constraint is obtained under some prior regarding the values of the NS spins~\cite{GW170817}. The best one probably corresponds to the low-spin scenario (say, with dimensionless spin parameter $\vert\chi\vert\leqslant 0.05$), since we expect from binary pulsar observations in our galaxy that the spin-orbit and spin-spin terms will make negligible contributions to the accumulated phase of NS binaries. The data analysis process should improve in this regard when we have more detections and higher signal-to-noise ratios, so that we can measure independently the NS spins using PN templates. The precise expression  of $\tilde{\Lambda}^{(2)}$ in terms of the companion parameters is~\cite{FH08, F14}\footnote{The normalization is chosen in such a way that $\tilde{\Lambda}^{(2)} = \Lambda^{(2)}_1 = \Lambda^{(2)}_2$ in the case of two identical neutron stars, \textit{i.e.}, with the same mass and the same EoS.}
\begin{equation}\label{eq:Lamdatilde}
\tilde{\Lambda}^{(2)} = \frac{16}{13}\,\Biggl[ \left( 1 + 11 \frac{m_2}{m}\right) \left(\frac{m_1}{m}\right)^4 \Lambda^{(2)}_1 + \left( 1 + 11 \frac{m_1}{m}\right) \left(\frac{m_2}{m}\right)^4 \Lambda^{(2)}_2 \Biggr]\,,
\end{equation}
where the individual dimensionless mass quadrupole deformability parameter is defined by
\begin{equation}\label{eq:Lamdadef}
\Lambda^{(2)}_A = \frac{2}{3}
k^{(2)}_A \biggl(\frac{R_A c^2}{G m_A}\biggr)^5\,.
\end{equation}
Here $m_A$ and $R_A$ are the mass and radius of the NS, whereas $k^{(2)}_A$ is a characteristic numerical coefficient called Love number~\cite{Love11}. Since Eq.~\eqref{eq:Lamdatilde} depends on both the masses and tidal deformabilities, with the masses being biased by gravitational lensing unlike the polarizabilities, the observable combination~\eqref{eq:Lamdatilde} could be used to recognize strongly lensed GW binary signals from unlensed ones with intrinsically higher masses~\cite{Pang:2020qow}. On the other hand, assuming the absence of lensing, the simultaneous measurement of $m_1$, $m_2$ and $\tilde{\Lambda}^{(2)}$ may provide an estimation of the redshift independently of electromagnetic observations~\cite{MessRead12}. This is particularly interesting for cosmography applications such as the measurement of the Hubble-Lema\^itre parameter.

Both the Love number and the compacity parameter $\mathcal{C}_A \equiv \frac{Gm_A}{R_A c^2}$ depend on the particular EoS. The Love numbers and tidal polarizabilities have been computed numerically for NS~\cite{Hind08, BinnP09, DN09tidal, HindLLR10}. Typically $k^{(2)} \simeq 0.1$ and $\mathcal{C} \simeq 0.15$,\footnote{By contrast, the Love numbers of black holes, \textit{i.e.}, in the limit where the compacity $\mathcal{C}_A\to \frac{1}{2}$, are exactly zero~\cite{FangLove05, BinnP09, DN09tidal, PhysRevD.92.024010, PhysRevD.91.104018}.} which means that the quadrupole deformation coefficient~\eqref{eq:Lamdadef} is a large number of the order of $\sim 1000$, hence the effect is large enough to be measurable from GW signals with a reasonable signal-to-noise ratio~\cite{FH08, DNV12, F14}. More precisely, we can estimate its magnitude by computing the tidal phase from Eq.~\eqref{eq:phaseSPA} at the point of contact of the two NS. At leading order for two identical NS (with common Love number $k^{(2)}$ and compacity $\mathcal{C}$), defining the contact point by $v_\text{contact}=\mathcal{C}^{1/2}$, we expect the maximal tidal phase lag to be roughly~\cite{DN09tidal}
\begin{equation}\label{eq:phiequalmass}
\psi^\text{max}_\text{tidal} = - \frac{39}{32} \,k^{(2)} \mathcal{C}^{-5/2} \simeq - 14~\text{rad}\,,
\end{equation}
which is amply sufficient for detection and data analysis (see \textit{e.g.}~\cite{3mn, CF94}).

The tidal polarizabilities~\eqref{eq:defpolarizability} are physical parameters to the extent that they directly parametrize the effective matter action~\eqref{Sm}--\eqref{Ltidal} we adopt in this work, following Refs.~\cite{BiniDF12, HFB20a}, as an efficient and elegant tool to describe tidal effects in the case of compact bodies. In the present paper, we shall analyze the tidal response of NS binaries and the modification of the GW phase to higher order, corresponding to mass quadrupole, current quadrupole and mass octupole tidal interactions. Accordingly, we introduce three tidal polarizability coefficients, conveniently denoted, using standard normalization~\cite{Hind08, HindLLR10}, as
\begin{subequations}\label{eq:defpolarizability}
\begin{align}
G \mu_{A}^{(2)} &\equiv \Bigl(\frac{G m_A}{c^{2}}\Bigr)^5\Lambda^{(2)}_A = \frac{2}{3} \,k_{A}^{(2)}
R_{A}^{5}\,,\\ G \sigma_{A}^{(2)} &=
\frac{1}{48} \,j_{A}^{(2)} R_{A}^{5}\,,\\ G \mu_{A}^{(3)} &= \frac{2}{15} \,k_{A}^{(3)} R_{A}^{7}\,,
\end{align}
\end{subequations}
and related to corresponding relativistic generalizations $k_{A}^{(2)}$, $j_{A}^{(2)}$ and $k_{A}^{(3)}$ of Love numbers for the mass quadrupole, current quadrupole and mass octupole moments of the body, with $R_A$ denoting its radius in a coordinate system such that the area of the sphere of radius $R_A$ is $4\pi R_A^2$.

As we have seen, the tidal mass quadrupole contribution to the phase~\eqref{eq:chirp} starts formally at 5PN order; here, we shall also compute the next-to-leading (NL) as well as the next-to-next-to-leading (NNL) corrections arising formally at 6PN and 7PN orders. The current quadrupole will start at NL/6PN order and we shall control the NNL/7PN term therein, while the mass octupole term will be purely a NNL/7PN contribution. We shall finally include the tidal contributions of GW tails to leading 6.5PN order, and to NL/7.5PN order.

In the formalism of effective action on which we lean, each compact object is described by an effective point particle endowed with internal structure. The effect of the internal structure is described by some non-minimal matter couplings to gravity introduced at the level of the action, involving relativistic tidal moments given by appropriate covariant derivatives of the Riemann tensor (or its dual), evaluated at the location of the particle, and partly contracted with several occurrences of the four-velocity vector. 

A regularization is required to removing the self field of the point-like object, thus ``automatically'' selecting the external tidal field experienced by the body $A$ due to the other bodies $B\not= A$ composing the system. We rely on dimensional regularization, which is known to give a complete physical answer in high PN approximations (see notably~\cite{DJSdim, BDE04}). However, up to the NNL order, as shown in \cref{app:proof}, it is equivalent to the simpler Hadamard ``partie finie'' regularization, so we actually use the Hadamard regularization in our practical calculations.

In the previous work~\cite{HFB20a}, we obtained the tidal effects in the conservative equations of motion (EoM) of compact binary systems to NNL order in the PN expansion. The internal structure and finite size of the compact objects were described by means of the Fokker action associated with the sum of the effective matter action and the Einstein-Hilbert gravitational action (with a gauge fixing term), through the three tidal polarizability coefficients~\eqref{eq:defpolarizability}. In particular, we obtained the invariant energy of the compact binary system in the case of quasi-circular orbits, which was found to be consistent with the (PN re-expansion) of the known effective-one-body (EoB) Hamiltonian~\cite{BiniDF12}.

In the present paper, we compute the tidal effects in the GW energy flux to NNL order within the so-called PN-matched Multipolar-post-Minkowskian (MPM-PN) formalism, which applies specifically in harmonic coordinates~\cite{BD86, BD88, BD89, BD92, B95, B98mult, PB02, BFN05}. The MPM-PN approach describes the waveform by means of mass and current radiative multipole moments defined in the asymptotic region, which are themselves related to some appropriate source-type multipole moments defined in the near zone for the whole matter system. Beware that we work with two different kinds of multipole moments in this article: the tidal moments, describing the individual deformation of the bodies, and the source multipole moments describing the mass distribution of the overall system. Note that, at the lowest order, the theory is linear, so that a given source multipole moment is the sum of its point-particle counterpart for the orbital motion and the corresponding tidal multipole moments of both bodies.

The energy flux computed in the present paper\footnote{All the results in this paper take into account the corrections of the equations described in the Erratum \cite{erratum}.}, together with the conservative energy deduced from the EoM in our previous work~\cite{HFB20a}, are the two basic ingredients required for insertion into the flux-balance equation and computation of the phase or frequency evolution to NNL order. As a matter of summary, we present here our end result for the tidal part of the SPA phase in the case of equal NS, with the same mass ($\nu=1/4$) and identical polarizabilities:\footnote{To present our results below we conveniently use a ``tilted'' notation for the polarizabily coefficients defined by Eqs.~\eqref{eq:polarpm} and~\eqref{eq:polarpmtilde}. For identical bodies, with Love numbers $k^{(2)}$, $j^{(2)}$ and $k^{(3)}$, and compacity $\mathcal{C}$, such notation reduces to (with \textit{e.g.} $\widetilde{\mu}^{(2)}\equiv \widetilde{\mu}_{+}^{(2)}$ and $\widetilde{\mu}_{-}^{(2)}=0$) 
$$\widetilde{\mu}^{(2)} = \frac{1}{48}\frac{k^{(2)}}{\mathcal{C}^{5}}\,,\qquad \widetilde{\sigma}^{(2)} = \frac{1}{1536}\frac{j^{(2)}}{\mathcal{C}^{5}}\,,\qquad \widetilde{\mu}^{(3)} = \frac{1}{960}\frac{k^{(3)}}{\mathcal{C}^{7}}\,.$$
The formula~\eqref{eq:phiequal} can easily be reconciled with Eq.~\eqref{eq:phaseSPA} to the leading order.}
\begin{align}\label{eq:phiequal}
\psi_\text{tidal} &= - \frac{117}{2} v^5 \biggl[ \widetilde{\mu}^{(2)} + \left( \frac{3115}{1248} \widetilde{\mu}^{(2)} + \frac{370}{117} \widetilde{\sigma}^{(2)} \right) v^{2} \nonumber\\&\qquad\qquad - \pi \widetilde{\mu}^{(2)} v^{3} + \left(\frac{29323235}{3429216} \widetilde{\mu}^{(2)} + \frac{935380}{66339} \widetilde{\sigma}^{(2)} + \frac{500}{351} \widetilde{\mu}^{(3)} \right) v^{4} - \pi \left(\frac{2137}{546} \widetilde{\mu}^{(2)} + \frac{592}{117} \widetilde{\sigma}^{(2)} \right)v^{5} \biggr]\,.
\end{align}
Note that, besides the even PN corrections, there also appear half-integer 6.5PN and 7.5PN contributions, which are due to propagating GW tails at infinity~\cite{BD88, BD92}. 

The rest of the paper is organized as follows. In \cref{sec:StressEnergyTensor}, we recall the matter action we start with (details of its construction are given in \textit{e.g.}~\cite{BiniDF12, HFB20a}) and compute the stress-energy tensor of the system as well as its 3+1 decomposition rewritten in a convenient form. Next, we calculate, in \cref{sec:potentials}, the potentials sourced by the previous stress-energy tensor (some long formulas are relegated to \cref{app:tidal}). In \cref{sec:MultipolarMoments} we apply the GW generation formalism, which yields the source multipole moments of the binary system in a general frame. Those are then specialized to the center of mass (CoM) frame and, in a last stage, for circular orbits (while the moments in a general frame are too long, we present the CoM moments in \cref{app:tidal}). As for the instantaneous GW flux, it is computed in \cref{sec:Flux} in a modal form based on the mode decomposition of the source multipole moments. The missing tail part is obtained from those flux modes.  In \cref{sec:Phase} we present our result for the phase evolution, both in the standard Taylor form and in the Fourier domain, using the SPA. Finally, in \cref{sec:conclusion}, we conclude and make comparisons with the existing literature. We prove in \cref{app:proof} that the dimensional and Hadamard regularizations are equivalent for this problem up to NNL order.

\section{Matter action and stress-energy tensor}\label{sec:StressEnergyTensor}

\subsection{General formalism}

In the preceding paper~\cite{HFB20a}, we analyzed the motion of a compact binary system including tidal interactions. To do so, we considered the gravitational Einstein-Hilbert action endowed with the standard harmonic gauge fixing term, to which we added the effective matter action for a system of $N$ massive gravitationally interacting compact bodies with internal structure. The motion was obtained by varying the associated Fokker action. The next crucial step in our approach, pursued in this section, consists in the computation of the matter stress-energy-tensor, whose vocation is to be inserted in a GW generation formalism. For that purpose, we need only the matter part of the action which admits the general form
\begin{equation}\label{Sm}
S_m = \sum_{A=1}^N \int \dd \tau_{A}\, L_A\,,
\end{equation}
where the term associated with particle $A$ integrates over its proper time variation $\dd\tau_A$ which is such that the four velocity $c u^{\mu}_{A} = \dd y^{\mu}_{A}/\dd\tau_{A}$ is normalized to $g^A_{\mu\nu} u_{A}^{\mu}u_{A}^{\mu} = -1$. Here, $g^A_{\mu\nu}$ means that the metric is evaluated at the location of the particle $A$, with the self-field contribution from $A$ removed with the help of an appropriate self-field regularization, namely dimensional regularization.

In the approximation of point particles (pp) deprived of internal structure and unresponsive to tidal fields, the action is given by the standard mass term. To describe the response of the internal structure of the compact objects to tidal interactions, we add to the point-particle action the following specific non-minimally coupled piece:\footnote{We use the same conventions and notation as in Ref.~\cite{HFB20a}. See~\cite{BiniDF12, HFB20a} for more details, as well as~\cite{ThH85, Zhang86, DSX1, FangLove05, FH08, BinnP09, DN09tidal, DN10} for preceding fundamental works and alternative discussions. See also~\cite{Dixon64, Dixon73} for general definitions of the Dixon moments, including spins, or~\cite{BIsrael75, SP10, M15} for a more practical approach at the level of the action.}
\begin{equation}\label{Ltidal}
L_A =  - m_A c^2 + \frac{\mu_{A}^{(2)}}{4}
G^{A}_{\mu\nu}G_{A}^{\mu\nu} +
\frac{\sigma_{A}^{(2)}}{6c^{2}}H^{A}_{\mu\nu}
H_{A}^{\mu\nu} +
\frac{\mu_{A}^{(3)}}{12} G^{A}_{\lambda\mu\nu}
G_{A}^{\lambda\mu\nu} \,.
\end{equation}
To the NNL order investigated in this paper, it is sufficient to consider the above three terms\footnote{Other multipoles, proper time derivatives of tidal moments, as well as cubic combinations of those quantities, can be checked to appear at higher post-Newtonian order.}, made of quadratic products of tidal mass and current multipole moments, namely the mass quadrupole tidal moment $G^{A}_{\mu\nu}$, the current quadrupole $H^{A}_{\mu\nu}$ and the mass octupole $G^{A}_{\lambda\mu\nu}$. They are defined as\footnote{The dual of the Riemann tensor $R_{\mu \nu \rho \sigma}$ is defined as $R^{*}_{\mu \nu \rho \sigma} \equiv \tfrac{1}{2}\varepsilon_{\mu \nu \lambda \kappa} \,R\indices{^\lambda^\kappa_\rho_\sigma}$, where $\varepsilon_{\mu \nu \lambda \kappa}$ stands for the totally anti-symmetric Levi-Civita tensor, with  $\varepsilon_{0123}=\sqrt{-g}$. The underlined indices are to be excluded from the operation of symmetrization.}
\begin{subequations}\label{eq:tidalR}
	\begin{align}
	G^A_{\mu\nu} &= - c^2 R^A_{\mu\rho\nu\sigma}
	u_A^{\rho}u_A^{\sigma}\,, \\ H^A_{\mu\nu} &= 2 c^3 
    R^{*\, A}_{(\mu\underline{\rho}\nu)\sigma}
  u_A^{\rho}u_A^{\sigma}\,, \\ G^A_{\lambda\mu\nu} &= - c^2 \nabla^\perp_{(\lambda}
	R^A_{\mu\underline{\rho}\nu)\sigma} u_A^{\rho}u_A^{\sigma}\,.
	\end{align}
\end{subequations}
The Riemann tensor and its dual are evaluated at point $A$ following the regularization, and we denote $\nabla^\perp_{\lambda} R^A_{\mu \nu \rho \sigma}\equiv(\nabla^\perp_{\lambda} R_{\mu \nu \rho \sigma})_A$ the projected covariant derivative, defined by $(\nabla^\perp_\lambda)_A=(\perp_\lambda^\kappa \nabla_\kappa)_A$ with $(\perp_\lambda^\nu)_A = (\delta_\lambda^\nu + u_\lambda u^\nu)_A$. The polarizability coefficients were already introduced in Eqs.~\eqref{eq:defpolarizability}.

The motivation for writing the Lagrangian~\eqref{Ltidal} stems from the fact that the matter action for a given body, in the limit of small radius relevant for compact objects, can be expanded near the worldline of a representative point on which the resulting action is then localized, and that, in the absence of spins, it can be only built from the metric and its derivatives in a way that preserves parity and general covariance. We already emphasized the crucial role played by the self-field regularization, which must properly be the dimensional regularization in this framework.

In order to compute the stress-energy tensor, we first shift from the action~\eqref{Sm} parametrized by the particle's proper time $\tau_A$ to an action defined in terms of an arbitrary parametrization $\bar{\tau}$. For instance, this parametrization can be the same for all particles. Once this is done, the ensuing expression for the action is manifestly invariant by reparametrization. We thus pose (with $\bar{u}_A^2 = g_{\mu\nu}^A\bar{u}_A^\mu\bar{u}_A^\nu$)
\begin{equation}
\dd\tau_A = \dd\bar{\tau}\sqrt{-\bar{u}_A^2}\,,\qquad u_A^\mu = \frac{\bar{u}_A^\mu}{\sqrt{-\bar{u}_A^2}}\,,\qquad L_A = \frac{\bar{L}_A}{\sqrt{-\bar{u}_A^2}}\,.
\end{equation}
The original action~\eqref{Sm} now becomes (for simplicity's sake, we suppress the particle's label till the end of this section)
\begin{equation}\label{eq:Sparaminv}
S_m = \sum \int \dd\bar{\tau} \,\bar{L}\Bigl(\bar{u}^{\mu}, g_{\mu\nu}, R_{\mu \nu \rho \sigma},\nabla_{\lambda}R_{\mu \nu \rho \sigma}\Bigr)\,.
\end{equation}
As it is written, the Lagrangian $\bar{L}$ is an (ordinary) function of independent variables: the arbitrary parametrized four-velocity $\bar{u}^{\mu}$, the covariant metric, the Riemann tensor and the covariant derivative of the Riemann tensor. The configuration variables are just the particle's positions $y^\mu(\bar{\tau})$ and their derivatives $\bar{u}^\mu(\bar{\tau})$. We thus define the linear momentum $p_\mu$ as the conjugate momentum of the position, \textit{i.e.},
\begin{equation}\label{eq:pmu}
p_{\mu} = \dfrac{\partial \bar{L}}{\partial \bar{u}^{\mu}}\,.
\end{equation}
Following Refs.~\cite{Dixon64, Dixon73, BIsrael75, M15}, we further introduce the quadrupole current $J^{\mu \nu \rho \sigma}$ and octupole current $J^{\lambda \mu \nu \rho \sigma}$ as\footnote{The chosen prefactors match previous definitions in the literature~\cite{M15}. As shown in the Appendix A of~\cite{M15}, they are such that $J^{\mu \nu \rho \sigma}$ and $J^{\lambda \mu \nu \rho \sigma}$ coincide with the Dixon quadrupole and octupole moments~\cite{Dixon64, Dixon73}, respectively, at the considered approximation level. We refer to~\eqref{eq:J} as multipole ``currents'' in order to reduce confusion with the tidal moments $G_L$, $H_L$ as well as with the source multipole moments $I_L$, $J_L$ considered in \cref{sec:MultipolarMoments}.}
\begin{equation}\label{eq:J}
J^{\mu \nu \rho \sigma} = - 6 \dfrac{\partial \bar{L}}{\partial R_{\mu \nu \rho \sigma}}\,,\qquad
J^{\lambda \mu \nu \rho \sigma} = - 12 \dfrac{\partial \bar{L}}{\partial \nabla_{\lambda}R_{\mu \nu \rho \sigma}}\,.
\end{equation}
The current $J^{\mu \nu \rho \sigma}$, and the current $J^{\lambda \mu \nu \rho \sigma}$ on its four last indices, have the same symmetries as the Riemann tensor. In addition, $J^{\lambda \mu \nu \rho \sigma}$ satisfies the cyclic symmetry $J^{[\lambda \mu \nu] \rho \sigma}=0$ as a consequence of the Bianchi identity.

By varying the action with respect to the worldline of the particle we obtain the EoM~\cite{Dixon73}
\begin{align}
\frac{\dD p_{\mu}}{\dd\tau} = -\frac{1}{6}J^{\nu \rho \sigma \kappa}\nabla_{\mu}R_{\nu \rho \sigma \kappa} -\frac{1}{12}J^{\lambda \nu \rho \sigma \kappa}\nabla_{\mu} \nabla_{\lambda} R_{\nu \rho \sigma \kappa}\,.
\end{align}
Next, the stress-energy tensor is obtained by variation with respect to the metric. With the action depending on the Riemann tensor and its first covariant derivative, we obtain it as the sum of pole, quadrupole and octupole pieces~\cite{M15},
\begin{equation}
T^{\mu \nu} = T^{\mu \nu}_{\text{pole}} + T^{\mu \nu}_{\text{quad}} + T^{\mu \nu}_{\text{oct}}\,.
\end{equation}
There is no dipole contribution since we neglect the spins. The pole part takes the usual form of the stress-energy tensor of a particle with worldline $y^\mu$, four-linear momentum $p_\mu$ and four-velocity $u^\mu$ (parametrized by $\tau$), namely
\begin{equation}\label{eq:Tpole}
T^{\mu \nu}_{\text{pole}} = \int \dd\tau \,p^{(\mu}u^{\nu)} \dfrac{\delta^{(4)}(x-y)}{\sqrt{-g}}\,,
\end{equation}
(with $\delta^{(4)}(x-y)$ the four-dimensional Dirac distribution), while the quadrupolar and octupolar pieces are given by
\begin{subequations}\label{eq:Tquadoct}
	\begin{align}
	T^{\mu \nu}_{\text{quad}} &= \int \dd\tau \biggl[ \dfrac{1}{3} R\indices{^(^\mu_\lambda_\rho_\sigma}J^{\nu)\lambda \rho \sigma}  \biggr]\dfrac{\delta^{(4)}(x-y)}{\sqrt{-g}} + \nabla_{\rho}\nabla_{\sigma} \int \dd \tau \biggl[ -\dfrac{2}{3}J^{\rho(\mu \nu) \sigma}\biggr]\dfrac{\delta^{(4)}(x-y)}{\sqrt{-g}}\,,\label{eq:Tquad}\\
	T^{\mu \nu}_{\text{oct}} &= \int \dd\tau \biggl[  \dfrac{1}{6}\nabla^{\lambda}R\indices{^(^\mu_\xi_\rho_\sigma}J\indices{_{\lambda}^\nu^)^\xi^\rho^\sigma} + \dfrac{1}{12}\nabla^{(\mu}R_{\xi\tau\rho\sigma}J^{\nu)\xi\tau\rho\sigma} \biggr]\dfrac{\delta^{(4)}(x-y)}{\sqrt{-g}}\nonumber\\&+ \nabla_{\rho}\int \dd\tau \biggl[ -\dfrac{1}{6}R\indices{^(^\mu_\xi_\lambda_\sigma}J^{\underline{\rho}\nu)\xi\lambda\sigma}-\dfrac{1}{3} R\indices{^(^\mu_\xi_\lambda_\sigma}J^{\nu)\rho\xi\lambda\sigma} + \dfrac{1}{3} R\indices{^\rho_\xi_\lambda_\sigma}J^{(\mu\nu)\xi\lambda\sigma}\biggr]\dfrac{\delta^{(4)}(x-y)}{\sqrt{-g}} \nonumber\\&+ \nabla_{\lambda}\nabla_{\rho}\nabla_{\sigma}\int \dd\tau \biggl[ \dfrac{1}{3} J^{\sigma\rho(\mu\nu)\lambda}\biggr]\dfrac{\delta^{(4)}(x-y)}{\sqrt{-g}}\,.\label{eq:Toct}
	\end{align}
\end{subequations}
As the latter formulas are general~\cite{Dixon64, Dixon73, BIsrael75, M15}, we can apply them to the specific case of the Lagrangian~\eqref{Ltidal}. For simplicity, we present the results setting $c=1$. Note the useful formula which links the current quadrupole invariant to the mass quadrupole invariant:
\begin{equation}
H_{\mu\nu}H^{\mu\nu} = 4 G_{\mu\nu}G^{\mu\nu} + 2 R_{\mu\nu\sigma\kappa}R\indices{^\mu^\nu^\sigma_\lambda}u^{\kappa}u^{\lambda}\,.
\end{equation}
The linear momentum is then found to be
\begin{align}\label{eq:pmuvalue}
	p_{\mu} &=  m\, u_\mu + \mu^{(2)}\left[ - R\indices{_\mu_\alpha_\gamma_\beta}u^{\gamma}G^{\alpha \beta} + \dfrac{3}{4} u_{\mu} G^{\alpha \beta}G_{\alpha \beta} \right] + \sigma^{(2)}\left[\dfrac{4}{3}R^*_{(\mu \underline{\alpha}\gamma) \beta}  u^{\gamma}H^{\alpha \beta}  + \dfrac{1}{2}H^{\alpha \beta}H_{\alpha \beta}u_{\mu}\right] \\
	& \qquad + \mu^{(3)}\left[ \frac{1}{4} G_{\alpha \beta \gamma}G^{\alpha \beta \gamma} u_{\mu} - \frac{1}{3} G^{\alpha \beta \gamma} \nabla^{\bot}_{\alpha} R_{\beta \mu \gamma \rho} u^{\rho} - \frac{1}{6} G^{\alpha \beta \gamma} u_{\alpha} \nabla^{\bot}_{\mu} R_{\beta \rho \gamma \sigma} u^{\rho}u^{\sigma} - \frac{1}{6} \bot_{\mu\alpha} G^{\alpha \beta \gamma} u^{\kappa}\nabla_{\kappa} R_{\beta \rho \gamma \sigma} u^{\rho}u^{\sigma}\right] \,,\nonumber
\end{align}
and we observe, as a check, the consequence of the invariance of the action by worldline reparametrization, namely 
\begin{equation}\label{eq:pmuunu}
p_\mu u^\mu =  - m + \frac{\mu^{(2)}}{4}
G_{\mu\nu}G^{\mu\nu} +
\frac{\sigma^{(2)}}{6}H_{\mu\nu}
H^{\mu\nu} +
\frac{\mu^{(3)}}{12} G_{\lambda\mu\nu}
G^{\lambda\mu\nu} = L \,.
\end{equation}
On the other hand, the explicit expressions of the quadrupole and octupole currents read
\begin{subequations}\label{eq:Jvalue}
	\begin{align}	
	J^{\mu \nu \rho \sigma} &= \mu^{(2)}\left( - 3 u^{[ \mu}G^{\nu ] [ \rho} u^{\sigma ] } \right) + \sigma^{(2)}\left( \varepsilon\indices{^\mu^\nu_\alpha_\beta}u^{\alpha}H^{\beta [\rho}u^{\sigma ]} + \varepsilon\indices{^\rho^\sigma_\alpha_\beta}u^{\alpha}H^{\beta [\mu}u^{\nu ]} \right)\,,\\
	J^{\lambda \mu \nu \rho \sigma} &= \mu^{(3)}\left( -2  \perp_\kappa^\lambda u^{[ \mu}G^{\nu ] \kappa [ \rho}u^{\sigma ]} \right)\,,
	\end{align}
\end{subequations}
thereby completing the dynamics. As a verification of the EoM and stress-energy tensor, we performed a direct variation of the mass quadrupole contribution $\propto \mu_{A}^{(2)}$ in the action~\eqref{Ltidal}, \textit{i.e.}, without using the general formalism~\eqref{eq:Sparaminv} nor the definitions of $p_{\mu}$, $J^{\mu \nu \rho \sigma}$ and $J^{\lambda \mu \nu \rho \sigma}$, which led us to an equivalent result.

\subsection{Ready-to-use expressions}
\label{sec:ready}

Equations~\eqref{eq:Tpole}--\eqref{eq:Tquadoct} express the matter stress-energy tensor, together with the explicit expressions~\eqref{eq:pmuvalue} of the linear momentum and~\eqref{eq:Jvalue} of the currents, in terms of the tidal multipole moments. In turn, the tidal moments are given in terms of the metric, curvature and matter variables by Eqs.~\eqref{eq:tidalR}. In this section and the next one, we need to rephrase the previous results in a more suitable way. The matter stress-energy tensor takes the general form
\begin{equation}\label{eq:TwithCD}
T^{\mu \nu} = \sum_{A}\left[ U^{\mu \nu}_{A} \delta_{A} + \nabla_{\alpha}\left(U_{A}^{\mu \nu \alpha} \delta_{A} \right) + \nabla_{\alpha}\nabla_{\beta}\left(U_{A}^{\mu \nu \alpha \beta} \delta_{A} \right) + \nabla_{\alpha} \nabla_{\beta} \nabla_{\gamma}\left(U_{A}^{\mu \nu \alpha \beta \gamma} \delta_{A} \right)\right]\,,
\end{equation}
where we use the coordinate-time $t$ parametrization and denote $\delta_A \equiv \delta^{(3)}[\mathbf{x}-\bm{y}_A(t)]$ the usual three-dimensional Dirac distribution, and where (for $u_A^0=\dd t/\dd\tau_A$)
\begin{subequations}
\begin{align}
U^{\mu \nu}_{A} &= \dfrac{1}{u_{A}^{0} \sqrt{-g}} \left(p_{A}^{(\mu}u_{A}^{\nu)} + \dfrac{1}{3} R\indices{_A^(^\mu_\lambda_\rho_\sigma}J_{A}^{\nu) \lambda\rho \sigma} + \dfrac{1}{6} \nabla_{\lambda} R\indices{_A^(^\mu_\xi_\rho_\sigma}J_{A}^{\underline{\lambda}\nu) \xi \rho \sigma} + \dfrac{1}{12} \nabla^{(\mu}R_{A \xi \tau \rho \sigma}J_{A}^{\nu) \xi \tau \rho \sigma}\right), \\
U_{A}^{\mu \nu \alpha} &= \dfrac{1}{3 u_{A}^{0} \sqrt{-g}} \left( - \dfrac{1}{2} R\indices{_A^(^\mu_\xi_\lambda_\sigma}J_{A}^{\underline{\alpha}\nu)\xi \lambda \sigma} - R\indices{_A^(^\mu_\xi_\lambda_\sigma}J_{A}^{\nu)\alpha \xi \lambda \sigma} + R\indices{_A^\alpha_\xi_\lambda_\sigma}J_{A}^{(\mu \nu) \xi \lambda \sigma} \right),\\
U_{A}^{\mu \nu \alpha \beta} &=  -\dfrac{2}{3 u_{A}^{0} \sqrt{-g}} J_{A}^{\alpha (\mu \nu) \beta},\\
U_{A}^{\mu \nu \alpha \beta \gamma} &= \dfrac{1}{3 u_{A}^{0} \sqrt{-g}} J_{A}^{\gamma \beta (\mu \nu) \alpha}.
\end{align}
\end{subequations}
Note that all the $U$'s are symmetric over $\mu$ and $\nu$; moreover, $U_{A}^{\mu \nu \alpha \beta}$ and $U_{A}^{\mu \nu \alpha \beta \gamma}$, over their 4 first indices, have the same symmetries as the Jacobi tensor $R^{\alpha (\mu \nu) \beta}$ [remind the definitions~\eqref{eq:J}].

By expliciting the covariant derivatives in~\eqref{eq:TwithCD} as the sum of partial derivatives and Christoffel symbols, we get some ready-to-use formulas that are directly entered into our computational codes,\footnote{All our calculations are done with the software \textit{Mathematica} and the tensor package \textit{xAct}~\cite{xtensor}.} namely
\begin{equation}\label{eq:TwithPD}
T^{\mu \nu} = \sum_{A}\left[ \mathcal{T}^{\mu \nu}_{\text{M}} + \dfrac{1}{\sqrt{-g}}\,\partial_{\alpha}\left(\mathcal{T}_{\text{D}}^{\mu \nu \alpha} \delta_{A} \right) + \dfrac{1}{\sqrt{-g}}\,\partial_{\alpha \beta}\left(\mathcal{T}_{\text{Q}}^{\mu \nu \alpha \beta} \delta_{A} \right) + \dfrac{1}{\sqrt{-g}}\,\partial_{\alpha \beta \gamma}\left(\mathcal{T}_{\text{O}}^{\mu \nu \alpha \beta \gamma} \delta_{A} \right)\right]\,,
\end{equation}
where
\begin{subequations}
\begin{align}
\mathcal{T}^{\mu \nu}_{\text{M}} &= U^{\mu \nu} + 2\Gamma^{(\mu}_{\lambda \rho}U^{\nu) \lambda \rho} + \left[\partial_\lambda \Gamma^{(\mu}_{\rho \sigma} + \Gamma^\kappa_{\rho \sigma}\Gamma^{(\mu}_{\lambda \kappa} \right] U^{\nu)\lambda \rho \sigma} - \Gamma^{(\mu}_{\lambda \kappa}\Gamma^{\nu)}_{\rho \sigma}U^{\kappa \lambda \rho \sigma} \\
&+ \left[ \partial_{\lambda} \Gamma^{(\mu}_{\gamma \delta} \,\Gamma^{\nu)}_{\rho \sigma} +\partial_{\sigma}\left( \Gamma^{(\mu}_{\rho \lambda} \Gamma^{\nu)}_{\gamma \delta}\right) - 2 \Gamma^{\kappa}_{\lambda \sigma}\Gamma^{(\mu}_{\kappa \rho}\Gamma^{\nu)}_{\gamma \delta} + \Gamma^{\kappa}_{\gamma \delta}\Gamma^{(\mu}_{\kappa \rho}\Gamma^{\nu)}_{\lambda \sigma} -2 \Gamma^{\kappa}_{\sigma \gamma}\Gamma^{(\mu}_{\rho \lambda}\Gamma^{\nu)}_{\kappa \delta} \right]U^{\rho\lambda \gamma \delta \sigma} \nonumber \\
&+ \left[ 2\Gamma^{\kappa}_{\sigma \gamma}\partial_{\lambda} \Gamma^{(\mu}_{\kappa \delta} +\Gamma^{\kappa}_{\sigma \lambda}\partial_{\kappa} \Gamma^{(\mu}_{\gamma \delta}  -\partial_{\sigma}\left(\Gamma^{\kappa}_{\gamma \delta} \Gamma^{(\mu}_{\kappa \lambda} \right) +\Gamma^{\rho}_{\gamma \delta}\Gamma^{\kappa}_{\sigma \lambda}\Gamma^{(\mu}_{\rho \kappa}+ 2\Gamma^{\rho}_{\kappa \delta}\Gamma^{\kappa}_{\sigma \gamma}\Gamma^{(\mu}_{\rho \lambda} -\partial_{\lambda\sigma} \Gamma^{(\mu}_{\gamma \delta} \right]U^{\nu)\lambda \gamma \delta \sigma} \, , \nonumber \\
\mathcal{T}_{\text{D}}^{\mu \nu \alpha} &= \sqrt{-g} \left[U^{\mu \nu \alpha} + \Gamma^\alpha_{\sigma \kappa}U^{\mu \nu \sigma \kappa} - 2 \Gamma^{(\mu}_{\sigma \kappa} U^{\nu) \alpha \sigma \kappa} -\partial_\sigma \Gamma^{\alpha}_{\rho \lambda} U^{\mu \nu \rho \lambda \sigma} + \partial_\sigma \Gamma^{(\mu}_{\rho \lambda} \left( 2U^{\nu) \alpha \rho \lambda \sigma} +U^{\nu)\sigma \rho \lambda \alpha}  \right) \right. \nonumber \\
& \left. \qquad \qquad + \Gamma^{(\mu}_{\lambda \gamma} \left(U^{\nu) \lambda \sigma\rho \alpha}\Gamma^{\gamma}_{\rho \sigma}-\Gamma^{\nu)}_{\rho \sigma}U^{\gamma \lambda \sigma \rho \alpha}  \right) +2\Gamma^{\rho}_{\sigma \kappa} \left( \Gamma^{\alpha}_{\rho \lambda}U^{\mu \nu \kappa \lambda \sigma}-\Gamma^{(\mu}_{\lambda \rho}U^{\nu)\alpha\kappa\lambda\sigma}  \right) \right. \nonumber \\
& \left. \qquad \qquad -2\Gamma^{(\mu}_{\rho \lambda} \left(U^{\nu) \alpha\kappa \lambda\sigma}\Gamma^{\rho}_{\sigma\kappa}-\Gamma^{\nu)}_{\sigma \kappa}U^{\alpha \kappa\rho\lambda\sigma}  \right) + 2\left(\Gamma^{\alpha}_{\rho \lambda}\Gamma^{(\mu}_{\sigma \kappa}-\Gamma^{\alpha}_{\sigma \kappa}\Gamma^{(\mu}_{\rho \lambda}\right)U^{\nu)\kappa\rho\lambda\sigma}  \right]\,,\\
\mathcal{T}_{\text{Q}}^{\mu \nu \alpha \beta} &= \sqrt{-g}\left[ U^{\mu \nu \alpha \beta} + 2 \Gamma^{(\mu}_{\sigma \kappa}\left( U^{\nu) \sigma \alpha \beta \kappa}- U^{\nu) \alpha \sigma \kappa \beta}\right) + \Gamma^{\alpha}_{\sigma \kappa}\left( 2 U^{\mu \nu \sigma \beta \kappa} + U^{\mu \nu \sigma \kappa \beta} \right)  \right]\,,\\
\mathcal{T}_{\text{O}}^{\mu \nu \alpha \beta \gamma} &= \sqrt{-g}U^{\mu \nu \alpha \beta \gamma}\,.
\end{align}
\end{subequations}

Finally, the basic matter variables that we use in our GW generation formalism are defined by
\begin{equation}\label{eq:sigma}
\sigma \equiv \frac{T^{00}+T^{ii}}{c^2}\,,\qquad \sigma_{i} \equiv \frac{T^{0i}}{c}\,, \qquad \sigma_{ij} \equiv T^{ij}\,.
\end{equation}
These quantities will comprise a point-particle part and a tidal part. The point-particle (pp) part is defined by the usual expression corresponding to the minimal coupling to the metric, \textit{i.e.}, Eq.~\eqref{eq:Tpole} in which $p_\mu$ is replaced by $m u_\mu$, the first term in~\eqref{eq:pmuvalue}, so that we have
\begin{subequations}\label{eq:pppart}

	\begin{align}
	\sigma_{\text{pp}} &= \frac{m_1 u_1^0}{\sqrt{-g_1}}\,\biggl(1+\frac{v_1^2}{c^2}\biggr)\delta_1 + 1 \leftrightarrow 2,\\ 
	(\sigma_i)_{\text{pp}} &= \frac{m_1 u_1^0}{\sqrt{-g_1}}\,v_1^i\,\delta_1 + 1 \leftrightarrow 2,\\
	(\sigma_{ij})_{\text{pp}} &= \frac{m_1 u_1^0}{\sqrt{-g_1}}\,v_1^i v_1^j\,\delta_1+ 1 \leftrightarrow 2\,,
	\end{align}
\end{subequations}
where $m_1$ is the constant PN mass, the three-dimensional Dirac distribution $\delta_1$ is confined to the worldline $y^i_1(t)$, $v_1^i=\dd y_1^i/\dd t$ or $v_1^\mu=(c, v_1^i)$ denote the ordinary coordinate velocity, $u_1^0=[-(g_{\mu\nu})_1 v_1^\mu v_1^\nu/c^2]^{-1/2}$ stands for the Lorentz factor, and $1 \leftrightarrow 2$ is the contribution of the other particle. Beware that the point-particle part~\eqref{eq:pppart} will actually involve ``indirectly'' tidal effects contained into the potentials parametrizing the metric as computed in Sec.~\ref{sec:potentials}.

In order to compute the multipole moments of the system $I_L$, $J_L$ defined in \cref{sec:MultipolarMoments}, we require $\sigma$ to be known at NNL order, $\sigma_{i}$ at NL and $\sigma_{ij}$ at leading order, for both the point-particle and tidal parts. For the treatment of the tidal corrections, it is convenient to split the temporal and spatial indices of \cref{eq:TwithPD}. We then obtain the complete, ready-to-use expressions for the ``direct'' tidal parts $\sigma_\text{tidal}$, $(\sigma_i)_\text{tidal}$ and $(\sigma_{ij})_\text{tidal}$ in terms of the tidal multipole moments; these are reported in \cref{app:tidal}.

The tidal moments $G_{ij}$, $H_{ij}$ and $G_{ijk}$ (when evaluated at point 1) have been computed in Eqs.~(4.1) of~\cite{HFB20a}. However, in order to present the expressions of $\sigma_\text{tidal}$, $(\sigma_i)_\text{tidal}$ and $(\sigma_{ij})_\text{tidal}$ as shown in Eqs.~\eqref{eq:sigmatidal} and everywhere henceforth, like for instance in \cref{eq:V1PN}, we rather use the tetradic components of these moments, denoted $\hatG_{ab}$, $\hatH_{ab}$ and $\hatG_{abc}$, obtained by projection on the worldline tetrad $e\indices{_\alpha^\mu} = (e\indices{_0^\mu}, e\indices{_a^\mu})$ constructed as follows: 
\begin{subequations}\label{eq:tetrad}
	\begin{align}
	e\indices{_0^\mu} &\equiv u^{\mu}\,,\\
	e\indices{_a^\mu} &= \left(\gamma^{\mu i}-\gamma^{\mu 0} \dfrac{v^i}{c}\right)e_{ai}\qquad\text{with}\qquad e_{ai} \equiv \left(\sqrt{\gamma}\right)_{ai}\,.
	\end{align}
\end{subequations}
Here $\gamma^{\mu \nu} = g^{\mu \lambda}\! \perp_\lambda^\nu$ is the inverse of the positive-definite metric $\gamma_{\mu\nu}=g_{\mu\nu}+u_\mu u_\nu$ induced on the hypersurface orthogonal to $u^\mu$ at the intersection point with the worldline, and the spatial tetrad vectors are defined from the square root  $(\sqrt{\gamma})_{ai}$ of the positive definite symmetric matrix $\gamma_{ij}$. One can show that this basis is complete and orthonormal (for more details, see~\cite{BFMP15}). 

Remembering that the tidal moments are defined in the particle's local frame orthogonal to the four velocity, \textit{i.e.} $\hatG_{0\alpha}=\hatH_{0\alpha}=\hatG_{0\alpha\beta}=0$ (see~\cite{HFB20a} for discussion), we have for instance (similarly for $H_{ij}$ and $G_{ijk}$)
\begin{subequations}\label{eq:projmoment}
	\begin{align}
	G_{ij} &= e\indices{^a_i} \,e\indices{^b_j} \,\hatG_{ab}\,,\\ 
	\hatG_{ab} &= \biggl[ e\indices{_a^i} \,e\indices{_b^j} - 2\frac{v_1^i}{c} \,e\indices{_{(a}^0} \,e\indices{_{b)}^j} + \frac{v_1^i v_1^j}{c^2} \,e\indices{_{a}^0} \,e\indices{_{b}^0} \biggr] G_{ij}\,,
	\end{align}
\end{subequations}
where $e\indices{^\beta_\nu}$ denotes the (transposed) inverse of $e\indices{_\alpha^\mu}$. The projection of the tidal tensors onto this tetrad simplifies significantly the computations, mostly because the projected three-dimensional tidal tensors become traceless. We know however, from the fact that the Lagrangian~\eqref{Ltidal} does not depend on the tetrad (see also the discussion in Sec.~II of Ref.~\cite{HFB20a}), that the final results are independent on a particular choice of tetrad $(e\indices{_0^\mu}, e\indices{_a^\mu})$ used in intermediate calculations. Other groups~\cite{VF13} may use different conventions for the tetrad with equivalent final results (see \cref{tab:literature} in \cref{sec:conclusion}).

\section{Computation of the metric potentials}\label{sec:potentials}

For this calculation, the metric, including tidal contributions, is required up to NNL 7PN order. This is in contrast with our previous work~\cite{HFB20a} on the NNL dynamics and EoM, where it was sufficient to insert the 2PN metric just for point particles, discarding internal structure effects; what made that possible were the specific properties of the Fokker action. Here, in order to get the multipole moments at the desired accuracy, we do need the 2PN metric including the tidal contributions therein. We employ our traditional parametrization by the set of elementary potentials $\{V,V_i,\hat{W}_{ij},\hat{X},\hat{R}_i \}$,\footnote{With a slight abuse of notation, the PN remainders $\mathcal{O}(c^{-8},c^{-7},c^{-6})$ means either that the metric is accurate to 2PN order in the standard sense, or that it is accurate to NNL order regarding tidal effects.}
\begin{subequations}\label{eq:metric}
\begin{align}
g_{00} &= -1 + \frac{2V}{c^{2}} - \frac{2V^{2}}{c^{4}}+ \frac{8}{c^{6}}
\left(\hat{X} + V_{i}V_{i} + \frac{V^{3}}{6} \right) +
\mathcal{O}\left( \frac{1}{c^{8}} \right)\,, \\ g_{0i} &=
-\frac{4V_{i}}{c^{3}} - \frac{8\hat{R}_{i}}{c^{5}} + \mathcal{O}\left(
\frac{1}{c^{7}} \right)\,, \\ g_{ij} &= \delta_{ij}\left(1 +
\frac{2V}{c^{2}} + \frac{2V^{2}}{c^{4}} \right) +
\frac{4\hat{W}_{ij}}{c^{4}} + \mathcal{O}\left( \frac{1}{c^{6}}
\right)\,.
\end{align}
\end{subequations}
The above full 2PN metric is used, after dropping all tidal terms, to compute the Riemann tensor and the tidal moments~\eqref{eq:tidalR}, which allows controlling the tidal parts~\eqref{eq:sigmatidal} of the matter currents. On the other hand, the NNL tidal effects in the metric~\eqref{eq:metric} are crucial for computing the point-particle parts of the matter currents~\eqref{eq:pppart} and the source multipole moments defined in~\cref{sec:MultipolarMoments}, which will be inserted later into the formula for the flux. Note that for the computation of the source multipole moments in \cref{eq:ILdecomp}, the only 2PN term is $\propto \sigma_\text{2PN}$, in which the metric only appears through $\sqrt{-g}$; in this calculation at NNL, $\hat{X}$ and $\hat{R}_i$ do not appear, meaning that only $V$ at NL order as well as $V_i$ and $\hat{W}_{ij}$ at leading order will be strictly necessary. These potentials are defined by
\begin{subequations}\label{eq:potentials}
\begin{align}
\Box V &= -4 \pi G \sigma\,, \\ 
\Box V_{i} &= -4  \pi G \sigma_{i}\,, \\ 
\Box \hat{W}_{ij} &= -4\pi G\bigl(\sigma_{ij} - \delta_{ij} \sigma_{kk} \bigr) -\partial_{i}V \partial_{j}V\,. 
\end{align}
\end{subequations}
The definitions are general and, of course, the source terms may involve both pp and tidal contributions, \textit{e.g.} $\sigma = \sigma_{\text{pp}} + \sigma_{\text{tidal}}$, where the tidal part  in terms of the tidal moments is displayed in Eqs.~\eqref{eq:sigmatidal}. 

The techniques we use for computing the potentials are well documented elsewhere (see \textit{e.g.}~\cite{BFP98, MHLMFB20}). In this work, dissipative radiation reaction effects can be ignored since they do not to contribute to the flux until the 2.5PN order, so that the Green function will be taken to be the symmetric one. As usual, it is essential to use a proper UV-type regularisation, namely dimensional regularization. In fact, we do not need for the present problem the corrections it brings with respect to simpler purely three-dimensional approaches, such as Hadamard's regularization, which gives equivalent results at the NNL tidal order. We present in \cref{app:proof} a detailed proof of this statement.

The tidal contributions to the metric obtained in present formalism show an interesting feature, already observed for binary systems of spinning compact objects in~\cite{BFMP15}: the tidal part of the potential $V$ contains a distributional term, which arises because of the distributional multi-derivatives in the expressions of the matter sources~\eqref{eq:sigmatidal}. To the lowest order, $\sigma_\text{tidal}$ is proportional to $\hatG\indices{_{1}_{ab}}\partial_{ab}(1/r_1)$, and, since $\partial_{ab}(1/r_1)=3\hat{n}_1^{ab}r_1^{-3}-\tfrac{4\pi}{3}\delta_{ab}\delta_1$, this leads to a distributional term for $V$ proportional to the trace $\delta^{ab}\hatG\indices{_1_{ab}}$, but which vanishes because the tidal tensors are projected onto the tetrad and are traceless. At the NL 6PN order, though, the distributional piece is non-zero and given by the Gel'fand-Shilov formula~\cite{MHLMFB20} as
\begin{equation}\label{eq:Vdistr}
V^{\text{distr}} = \frac{2 \pi}{5}\frac{G\mu_1^{(2)}}{c^2}\hatG\indices{_{1}_{ab}}v_1^a v_1^b \,\delta_1 + 1 \leftrightarrow 2 + \mathcal{O}\left( \dfrac{1}{c^{4}} \right).
\end{equation}
This term will not contribute to our calculation because the NL potential $V$ is only needed in a surface term at infinity where the UV regularization is irrelevant. However, it would be important to take into account if we were to evaluate the equivalent volume integral. For the ordinary part of the complete potential $V$ at the NL order, computed with Hadamard's regularization, we find 
\begin{align}\label{eq:V1PN}
V &= \dfrac{G m_{1}}{r_{1}} + \dfrac{3G \mu_{1}^{(2)}\hatG\indices{_{1}_{ab}}n_{1}^{a}n_{1}^{b}}{2r_{1}^{3}} \nonumber\\
& + \dfrac{1}{c^{2}} \Biggl\{ G m_{1}\left[-\dfrac{(n_{1}v_{1})^{2}}{2r_{1}} + \dfrac{2v_{1}^{2}}{r_{1}}+G m_{2}\left( -\dfrac{r_{1}}{4r_{12}^{3}} - \dfrac{5}{4r_{1}r_{12}} + \dfrac{r_{2}^{2}}{4r_{1}r_{12}^{3}}  \right)  \right] \nonumber\\
& \left. \qquad + \mu_{1}^{(2)}\left[ G \left( \dfrac{3}{4r_{1}}\hatG\indices{_{1}_{ab}}\hatG\indices{_{1}_{ab}} + \left( 3v_{1}^{2}- \dfrac{15}{4}(n_{1}v_{1})^{2} \right) \dfrac{\hatG\indices{_{1}_{ab}}n_{1}^{a}n_{1}^{b}}{r_{1}^{3}} + \dfrac{3}{2}(n_{1}v_{1})\left(\dfrac{\hatG\indices{_{1}_{ab}}n_{1}^{a}v_{1}^{b}}{r_{1}^{3}} - \dfrac{n_{1}^{a}n_{1}^{b}\partial_{t}\hatG\indices{_{1}_{ab}}}{r_{1}^{2}} \right) \right. \right. \right. \nonumber\\
& \qquad \qquad \qquad \left. \left. \left. + \dfrac{2n_{1}^{a}v_{1}^{b}\partial_{t}\hatG\indices{_{1}_{ab}}}{r_{1}^{2}} - \dfrac{n_{1}^{a}n_{1}^{b}\partial_{t}^{2}\hatG\indices{_{1}_{ab}}}{4r_{1}} \right) \right. \right. \nonumber\\
& \qquad \qquad \left. \left. + \dfrac{G^{2}m_{2}}{r_{12}^{3}} \left(\dfrac{3r_{12}}{2r_{1}^{2}}\hatG\indices{_{1}_{ab}}n_{12}^{a}n_{1}^{b} + \left( -\dfrac{6}{r_{1}}-\dfrac{3}{2r_{2}} \right) \hatG\indices{_{1}_{ab}}n_{12}^{a}n_{12}^{b} + \left(-\dfrac{3}{8r_{1}} - \dfrac{39r_{12}^{2}}{8r_{1}^{3}} + \dfrac{3r_{2}^{2}}{8r_{1}^{3}} \right)\hatG\indices{_{1}_{ab}}n_{1}^{a}n_{1}^{b} \right) \right. \right. \nonumber\\
& \qquad \qquad \left. + \dfrac{9}{2}\dfrac{G^{3}m_{2}^{2}}{r_{12}^{7}}\left((n_{12}n_{1})\left(1-\dfrac{r_1}{r_2}  \right) -\dfrac{r_{12}}{r_2} \right)\right] + \dfrac{4 G \sigma_{1}^{(2)}\varepsilon_{bij}\hatH\indices{_{1}_{aj}}n_{1}^{a}n_{1}^{b}v_{1}^{i}}{r_{1}^{3}} \Biggr\} + 1 \leftrightarrow 2 + \mathcal{O}\left( \dfrac{1}{c^{3}} \right)\,,
\end{align}
where we recall the definition~\eqref{eq:projmoment} of the projected tidal moments. Consistently with the approximation, we also included the ordinary point particle part at 1PN order. Using the same method, we computed $V_i$ at leading order,
\begin{equation}\label{eq:ViN}
V_{i} = \dfrac{G m_{1}}{r_{1}}v_{1}^{i} + \dfrac{3 G \mu_{1}^{(2)}\hatG\indices{_{1}_{ab}}n_{1}^{a}n_{1}^{b}}{2 r_{1}^{3}}v_{1}^{i} + \dfrac{G \mu_{1}^{(2)}n_{1}^{a}\partial_{t}\hatG\indices{_{1}_{ai}}}{2r_{1}^{2}} + \dfrac{G \sigma_{1}^{(2)}\varepsilon_{iab}\hatH\indices{_{1}_{ak}}n_{1}^{b}n_{1}^{k}}{r_{1}^{3}} + 1 \leftrightarrow 2 + \mathcal{O}\left( \dfrac{1}{c} \right).
\end{equation}
Notice that, due to the way the leading term of $(\sigma_i)_\text{tidal}$ is written in \cref{eq:sigmatidali}, some non-zero distributional terms are generated by multi-derivatives, but they cancel out in the end, so the potential $V_i$ does not contain any. For the potential $\hat{W}_{ij}$ at leading order, we get
\begin{align}\label{eq:Wij}
\hat{W}_{ij} &= \dfrac{G m_1}{r_1} \bigl( v_{1}^i v_{1}^j - \delta^{ij}v_1^2\bigr) + \dfrac{G^2 m_{1}^2}{4r_{1}^2}\bigl(n_{1}^{i}n_{1}^{j}-\delta_{ij}\bigr) - G^2 m_1 m_2 \partial_{1(i}\partial_{2j)}\ln S \nonumber \\
& + \mu_{1}^{(2)}\Biggl[\dfrac{G^2 m_2}{r_1 r_{12}^3} \left( \hatG\indices{_{1}_{ij}} -3 n_{12}^{(i}\hatG\indices{_{1}_{j)a}}n_{12}^{a} + 3 \delta^{ij} \hatG\indices{_{1}_{ab}}n_{12}^{a}n_{12}^{b} \right) - G^2m_2 \hatG\indices{_1_{ab}} \partial_{2(i}\partial_{1j)ab}\ln S \nonumber \\
& \qquad\quad + \dfrac{G}{r_{1}^3}\left( \dfrac{r_{1}^2}{2} \partial_{t}^2 \hatG\indices{_{1}_{ij}} +r_1 n_{1}^a v_{1}^{(i}\partial_t \hatG\indices{_{1}_{j)a}} + \dfrac{3}{2} \hatG\indices{_{1}_{ab}}n_{1}^a n_{1}^b \left(v_{1}^i v_{1}^j - \delta^{ij}v_1^2 \right) -\delta^{ij} r_1 n_{1}^a v_{1}^b \partial_t \hatG\indices{_{1}_{ab}}\right) \Biggr] \nonumber\\
& + \dfrac{2G\sigma_{1}^{(2)}}{r_{1}^3} \biggl[\varepsilon\indices{^(^i_a_b}\left( v_{1}^{j)} \hatH\indices{_{1}_{ak}}n_{1}^b n_{1}^k  + \dfrac{r_1}{3}\partial_t \hatH\indices{_{1}_{j)a}}n_{1}^{b}\right) -\delta^{ij} \varepsilon_{abk}\hatH\indices{_{1}_{k l}}n_{1}^a n_{1}^l v_{1}^b \biggr]
\nonumber \\
& -G^2 \mu_{1}^{(2)}m_1 \hatG\indices{_1_{ab}} \biggl[\dfrac{5}{128}\partial_{ijab}\left[\ln\left(\dfrac{r_1}{r_0}\right) \right] + \dfrac{5}{16} \dfrac{n_{1}^a n_{1}^b}{r_{1}^4}\delta_{ij} + \dfrac{1}{4 r_{1}^4}\delta^{a(i}\left(n_{1}^{j)}n_{1}^b -\dfrac{3}{8} \delta^{j)b}\right)  \biggr] + 1 \leftrightarrow 2 + \mathcal{O}\left( \dfrac{1}{c} \right)\,.
\end{align}
The point-particle part is depicted in the first line, where we denote $\partial_{Ai} \equiv \partial/\partial y_{A}^i$ and $S \equiv r_{1} + r_{2} + r_{12}$. These potentials satisfy $\Box V = \mathcal{O}(1/c^4)$, $\Delta V_i = \mathcal{O}(1/c^2)$ and $\Delta \hat{W}_{ij} = -\partial_i V \partial_j V+ \mathcal{O}(1/c^2)$ outside the particles. We also checked that they obey the harmonic gauge constraints
\begin{subequations}\label{eq:conteq}
\begin{align}
&\partial_t \left\{V + \dfrac{1}{c^2}\left[\dfrac{1}{2}\hat{W}+2V^2 \right] \right\} + \partial_i \left\{ V_i + \dfrac{2}{c^2}\left[\hat{R}_i + VV_i\right] \right\} = \mathcal{O} \left( \dfrac{1}{c^3} \right)\,, \\
& \partial_t V_i + \partial_j \left\{\hat{W}_{ij}- \dfrac{1}{2}\delta_{ij}\hat{W} \right\} = \mathcal{O} \left( \dfrac{1}{c} \right)\,,
\end{align}
\end{subequations}
which yield at the NL order the same EoM as obtained in Ref.~\cite{HFB20a}. This test confirms the values of all potentials that are required for the integration of the source multipole moments in \cref{sec:MultipolarMoments}. Note that, for this verification, we had to determine $V_{i}$ at NL and also $\hat{R}_{i}$ at lowest order, where $\hat{R}_{i}$ is defined by
\begin{equation}
\Box \hat{R}_{i} = -4 \pi G \bigl(V \sigma_{i} - V_{i} \sigma \bigr) -2
  \partial_{k}V \partial_{i}V_{k} - \frac{3}{2}\partial_{t}V
  \partial_{i}V\,.
\end{equation}
We do not give their values since they do not enter our later calculations.

\section{Computation of source multipole moments}\label{sec:MultipolarMoments}

The symmetric-trace-free (STF) multipole moments of isolated PN radiative sources are known from a matching between the inner PN expansion in the system near zone and the outer MPM expansion in the far zone~\cite{B98mult, PB02}. For any $\ell\geqslant 2$, they read
\begin{subequations}\label{eq:genMultMom}
\begin{align}
I_{L}(t)&= \underset{B=0}{\mathrm{FP}} \int \dd^{3}\mathbf{x} \left(\frac{r}{r_0}\right)^B \int_{-1}^{1}\dd z \left[ \delta_{\ell} \,\hat{x}_{L} \Sigma - \dfrac{4(2\ell+1) \delta_{\ell+1}}{c^2(\ell+1)(2\ell+3)} \hat{x}_{iL} \Sigma^{(1)}_{i} \right. \nonumber \\ &\left. \qquad \qquad \qquad \qquad \qquad \qquad  + \dfrac{2(2\ell+1)\delta_{\ell+2}}{c^{4}(\ell+1)(\ell+2)(2\ell+5)}  \hat{x}_{ijL} \Sigma^{(2)}_{ij} \right](\mathbf{x},t+zr/c)\,,\label{eq:IL}\\
J_{L}(t) &= \underset{B=0}{\mathrm{FP}} \int \dd^{3}\mathbf{x} \left(\frac{r}{r_0}\right)^B \int_{-1}^{1}\dd z \,\varepsilon_{ab \langle i_{\ell}} \left[ \delta_{\ell} \,\hat{x}_{L-1 \rangle a} \Sigma_{b} - \dfrac{2\ell+1}{c^2(\ell+2)(2\ell+3)} \delta_{\ell+1} \hat{x}_{L-1 \rangle ac} \Sigma^{(1)}_{bc} \right] (\mathbf{x},t+zr/c)\,.\label{eq:JL}
\end{align}
\end{subequations}
Here, $\hat{x}_L\equiv\text{STF}(x_{i_1}x_{i_2}\cdots x_{i_\ell})$ is the multipolar factor, the brackets surrounding indices refer to the STF projection, and the $\Sigma$'s (or their partial time-derivatives $\Sigma^{(n)}$'s), which must be evaluated at position $\mathbf{x}$ and at time $t+z \vert{\mathbf{x}}\vert/c$, are defined in terms of the PN expansion of the stress-energy pseudo-tensor $\bar{\tau}^{\mu\nu}$ in harmonic coordinates by
\begin{align}\label{eq:Sigma}
\Sigma \equiv \frac{\bar{\tau}^{00}+\bar{\tau}^{ii}}{c^2}\,,\qquad 
\Sigma_i \equiv \frac{\bar{\tau}^{0i}}{c}\,,\qquad
\Sigma_{ij} \equiv \bar{\tau}^{ij}\,.
\end{align}
The overbar refers to the PN expansion (see Sec.~II in~\cite{MHLMFB20} for further discussion). \cref{eq:sigma} gives the corresponding matter parts. The expressions of the source moments~\eqref{eq:genMultMom} are formally valid up to any PN order. In practice, their PN-expanded expressions are to be computed by means of the infinite PN series\footnote{The function $\delta_\ell (z)$ is defined [with $\delta(z)$ denoting the one-dimensional Dirac distribution] by
$$\delta_\ell (z) \equiv \frac{(2\ell+1)!!}{2^{\ell+1} \ell!}
	\,(1-z^2)^\ell\,, \quad \text{so that} \quad \int^1_{-1} \dd z\,\delta_\ell
	(z) = 1\,\quad \text{and} \quad \lim_{\ell \rightarrow +\infty}\delta_\ell (z) = \delta (z)\,.$$}
\begin{equation}\label{eq:intdeltal}
\int^1_{-1} \dd z~ \delta_\ell(z) \,\Sigma(\mathbf{x},t+z
	r/c) = \sum_{k=0}^{+\infty}\,\frac{(2\ell+1)!!}{(2k)!!(2\ell+2k+1)!!}
	\,\left(\frac{r}{c}\right)^{2k} \!\Sigma^{(2k)}(\mathbf{x},t)\,.
\end{equation}
An important feature of Eqs.~\eqref{eq:genMultMom} is the presence of the finite part ($\mathrm{FP}$) operation when some complex parameter $B$ goes to zero. The role of the finite part is to deal with the infra-red (IR) divergences initially introduced into the multipole moments by the fact that their PN-expanded integrands diverge at spatial infinity (as $r\rightarrow +\infty$). See Ref.~\cite{MHLMFB20} for details on how we deal in practice with this IR regularization. At the NNL order, we shall explicitly verify that the IR constant $r_0$ in Eqs.~\eqref{eq:genMultMom} never appears.

Like in previous works~\cite{BI04mult, MHLMFB20}, we find it convenient to decompose $I_L$ into three pieces corresponding to the three terms entering~\eqref{eq:IL}, referred to as scalar (S), vector (V) and tensor (T) terms. Applying the formula~\eqref{eq:intdeltal}, we further split each of these pieces into parts labelled I, II, III, $\dots$ according to their PN order. This leads to the decomposition of the $\ell$-th order mass-type moment to NNL order (omitting the PN remainders) as
\begin{subequations}\label{eq:ILdecomp}
\begin{align}
I_{L} &= \text{SI}_{L} + \text{SII}_{L} + \text{SIII}_{L} + \text{VI}_{L} + \text{VII}_{L} +\text{TI}_{L}\,,\\
\text{SI}_{L} &= \mathrm{FP} \int \dd^{3}\mathbf{x}  \,\hat{x}_{L} \left\{ \sigma + \dfrac{4 V}{c^{4}}\sigma_{ii} - \dfrac{2}{\pi G c^{4}}V_{i}\partial_{t} \partial_{i}V  - \dfrac{1}{\pi G c^{4}} \hat{W}_{ij}\partial_{ij}^{2}V - \dfrac{1}{2 \pi G c^{4}}(\partial_{t}V)^{2} + \dfrac{2}{\pi G c^{4}} \partial_{i}V_{j}\partial_{j}V_{i} \right. \nonumber\\
& \left. \hspace{4cm} - \dfrac{1}{2 \pi G c^{2}} \Delta(V^{2}) - \dfrac{2}{3 \pi G c^{4}}\Delta(V^{3}) - \dfrac{1}{2 \pi G c^{4}}\Delta(V \hat{W})\right\} \,,\\
\text{SII}_{L} &= \dfrac{1}{2c^{2}(2\ell +3)} \mathrm{FP} \dfrac{\dd^{2}}{\dd t^{2}}\int \dd^{3}\mathbf{x} \, \hat{x}_{L}  r^2\left\{ \sigma + \dfrac{4}{c^2}\sigma V - \dfrac{1}{\pi G c^{2}}\partial_i V \partial_i V \right\} \,,\\
\text{SIII}_{L} &= \dfrac{1}{8c^{4}(2\ell +3)(2\ell +5)} \mathrm{FP} \dfrac{\dd^{4}}{\dd t^{4}}\int \dd^{3}\mathbf{x}  \,\hat{x}_{L} r^{4} \sigma \,,\\
\text{VI}_{L} &= -\dfrac{4(2 \ell +1)}{c^{2}(\ell +1)(2\ell +3)} \mathrm{FP} \dfrac{\dd}{\dd t}\int \dd^{3}\mathbf{x}  \,\hat{x}_{iL} \left\{ \sigma_{i} +\dfrac{2}{c^{2}}\sigma_{i}V -\dfrac{2}{c^{2}}\sigma V_{i} + \dfrac{1}{\pi G c^{2}} \partial_{j}V \partial_{i}V_{j} \right. \nonumber\\
& \left. \qquad\qquad\qquad\qquad + \dfrac{3}{4 \pi G c^{2}}\partial_{t}V \partial_{i}V - \dfrac{1}{2 \pi G c^{2}}\Delta(V V_{i}) \right\} \,,\\
\text{VII}_{L} &= -\dfrac{2(2\ell+1)}{c^{4}(\ell+1)(2\ell +3)(2\ell +5)} \mathrm{FP} \dfrac{\dd^{3}}{\dd t^{3}}\int \dd^{3}\mathbf{x}  \,\hat{x}_{iL} \,r^{2} \sigma_{i} \,,\\
\text{TI}_{L} &= \dfrac{2(2\ell+1)}{c^{4}(\ell+1)(\ell +2)(2\ell +5)} \mathrm{FP} \dfrac{\dd^{2}}{\dd t^{2}}\int \dd^{3}\mathbf{x}  \,\hat{x}_{ijL} \left\{ \sigma_{ij} + \dfrac{1}{4 \pi G} \partial_{i}V\partial_{j}V \right\} \,.
\end{align}
\end{subequations}
Similarly, for the $\ell$-th order current moments to NNL order,
\begin{subequations}\label{eq:JLdecomp}
\begin{align}
J_{L} &= \text{VI}_{L} + \text{VII}_{L} + \text{TI}_{L}\,,\\
\text{VI}_{L} &= \varepsilon_{ab\langle i_{\ell}} \,\mathrm{FP}\int \dd^{3}\mathbf{x}  \,\hat{x}_{L-1 \rangle a} \left\{ \sigma_{b} + \dfrac{1}{c^{2}} \left[ 2 \bigl(\sigma_{b}V - \sigma V_{b}\bigr) + \dfrac{1}{\pi G} \left( \partial_{i}V \partial_{b}V_{i} + \dfrac{3}{4} \partial_{t}V \partial_{b}V - \dfrac{1}{2} \Delta(V V_{b})  \right) \right] \right\} \,\\
\text{VII}_{L} &= \dfrac{1}{2 c^{2}(2\ell +3)} \varepsilon_{ab \langle i_{\ell}} \mathrm{FP} \dfrac{\dd^{2}}{\dd t^{2}}\int \dd^{3}\mathbf{x}  \,r^{2} \hat{x}_{L-1\rangle a} \sigma_{b} \,,\\
\text{TI}_{L} &= -\dfrac{(2\ell+1)}{c^{2}(\ell +2)(2\ell+3)}\varepsilon_{ab\langle i_{\ell}} \mathrm{FP}\dfrac{\dd}{\dd t} \int \dd^{3}\mathbf{x}  \,\hat{x}_{L-1 \rangle ac} \left\{ \sigma_{bc} + \dfrac{1}{4 \pi G} \partial_{b}V \partial_{c}V \right\}\,.
\end{align}
\end{subequations}

The various encountered terms are of three types: (i) the compact support (C) terms, whose integrands are proportional to the matter currents $\sigma$'s, (ii) the non-compact (NC) support terms, whose volume integrals extend up to infinity, and (iii) the ``surface'' terms, also non-compact, but whose integrands are either pure divergences, or products of $\hat{x}_L$ and pure Laplacians. By integrating the latter terms by parts (taking into account the regularization factor $r^B$), one can transform them into easy-to-compute surface integrals (see Sec.~III C in~\cite{MHLMFB20} for details). In particular, assuming that the expansion of $F$ when $r\to\infty$ is power-like (without logarithms), it can be proved that
\begin{equation}\label{eq:LaplaceF}
\underset{B=0}{\mathrm{FP}} \int \dd^{3}\mathbf{x} \left(\frac{r}{r_0}\right)^B \hat{x}_L \,\Delta F = -4 \pi (2\ell+1) \bigl( F r^{\ell+1} \hat{n}_L \bigr)_\infty\,,
\end{equation}
where the notation $(\cdots)_\infty$ means the Hadamard partie finie regularization at infinity. With this formula, we have shown that, at NNL order, all the terms of this type for $F=\{V^2,V^3,V \hat{W}, VV_i \}$ vanish. The remaining terms (C and NC) can be integrated exactly. In the first case, we use the characteristic property of the Dirac distribution in the context of Hadamard's regularization, $\int \dd^3\mathbf{x}\, F(\mathbf{x},t) \,\delta_1=(F)_1$, where $(F)_1$ is the regularized value of the function $F$ at point $\mathbf{x}=\mathbf{y}_1$. There is a similar formula for delta's derivatives obtained by integration by parts. In the second case, we perform a brute-force integration after an appropriate change of variable as described in Sec.~V~D~3 of Ref.~\cite{BFeom}.

The explicit expressions of the (tidal parts of the) multipole moments of the system to NNL order are too long to be listed. However, they are substantially shortened by going to the frame of the center of mass (CoM). The conditions for going from a general frame to the CoM frame have been investigated in Sec.~V of Ref.~\cite{HFB20a}. For quantities in the CoM frame, it is convenient to redefine the polarizability parameters as
\begin{equation}\label{eq:polarpm}
\mu_\pm^{(\ell)} = \frac{1}{2}\left(\frac{m_{2}}{m_{1}}\,\mu_{1}^{(\ell)} \pm \frac{m_{1}}{m_{2}}\,\mu_{2}^{(\ell)}\right)\,,\qquad \sigma_\pm^{(\ell)} =
\frac{1}{2}\left(\frac{m_{2}}{m_{1}}\,\sigma_{1}^{(\ell)} \pm
\frac{m_{1}}{m_{2}}\,\sigma_{2}^{(\ell)}\right)\,,
\end{equation}
so that, for instance, $\mu_+^{(\ell)} = \mu_1^{(\ell)} = \mu_2^{(\ell)}$ and $\mu_-^{(\ell)} = 0$ when the two bodies are identical (with the same mass and internal structure). The tidal parts of the multipole moments in the CoM frame are reported in \cref{app:tidal}.

Next, we reduce the CoM moments for quasi-circular orbits following Sec. VI of~\cite{HFB20a}. To present the results, we introduce the normalized mass difference $\Delta \equiv \tfrac{m_1 - m_2}{m}$ and the PN parameter $\gamma=\frac{G m}{r c^2}$. We denote $\boldsymbol{n}$ the unit direction pointing from body 2 to 1, $\boldsymbol{\lambda}$ the unit vector perpendicular to $\boldsymbol{n}$ in the orbital plane, and $\boldsymbol{\ell}$ the unit vector perpendicular to the orbital plane, such that $(\boldsymbol{n},\boldsymbol{\lambda},\boldsymbol{\ell})$ forms a direct orthonormal triad. This implies notably that $\lambda^i=\frac{v^i}{r \omega}$ for exactly circular orbits, with $v^i=v^i_1-v^i_2$ representing the relative velocity. It is also convenient to use the adimensionalized versions of the polarizabilities~\eqref{eq:polarpm} (with $m=m_1+m_2$ representing the total mass)
\begin{equation}\label{eq:polarpmtilde}
\widetilde{\mu}_\pm^{(\ell)} = \left(\frac{c^2}{G m}\right)^{2\ell+1}
\!\!\!G\,\mu_\pm^{(\ell)}\,,\qquad \widetilde{\sigma}_\pm^{(\ell)} =
\left(\frac{c^2}{G m}\right)^{2\ell+1} \!\!\!G\,\sigma_\pm^{(\ell)}\,.
\end{equation}
The source moments for circular orbits including both the point-particle part (see~\cite{BFIS08}) and the tidal part then read
\begin{subequations}\label{eq:momentcirc}
\begin{align}
	& I_{ij}= m r^2 \Biggl[
	n^{\langle i} n^{j\rangle} \biggl\{\nu \biggl [1
	+ \Bigl(- \frac{1}{42}
	-  \frac{13}{14} \nu \Bigr) \gamma
	+ \Bigl(- \frac{461}{1512}
	-  \frac{18395}{1512} \nu  -  \frac{241}{1512} \nu^2\Bigr) \gamma^2\biggl]\nonumber\\
	& \qquad
	+ \Bigl(3 \widetilde{\mu}_{+}^{(2)}
	+ 3 \Delta\, \widetilde{\mu}_{-}^{(2)}\Bigr) \gamma^5
	+ \biggl [\widetilde{\mu}_{+}^{(2)} \Bigl(- \frac{3}{2}
	+ \frac{\nu}{7} -\frac{222}{7} \nu^2\Bigr)
	+ \Delta\, \widetilde{\mu}_{-}^{(2)} \Bigl(- \frac{3}{2}
	-  \frac{67}{7} \nu \Bigr)
	+ \frac{160}{3} \nu \widetilde{\sigma}_{+}^{(2)}\biggl] \gamma^6\nonumber\\
	& \qquad + \biggl [\widetilde{\mu}_{+}^{(2)} \Bigl(\frac{871}{56} - \frac{1613}{168} \nu - \frac{17237}{168} \nu^2
	+ \frac{929}{42} \nu^3\Bigr)
	+ \Delta\, \widetilde{\mu}_{-}^{(2)} \Bigl(\frac{871}{56}
	+ \frac{1493}{24} \nu - \frac{7201}{168} \nu^2\Bigr)
	+ \widetilde{\sigma}_{+}^{(2)} \Bigl(\frac{388}{9} \nu
	-  \frac{2504}{7} \nu^2\Bigr)\nonumber\\
	& \qquad + \frac{1732}{63} \Delta \nu \widetilde{\sigma}_{-}^{(2)}\biggl]
	\gamma^7\biggl\}
	+ \lambda^{\langle i} \lambda^{j\rangle} \biggl\{\nu \biggl [\Bigl(\frac{11}{21} -  \frac{11}{7} \nu \Bigr) \gamma
	+ \Bigl(\frac{1013}{378}
	+ \frac{299}{378} \nu
	-  \frac{365}{378} \nu^2\Bigr) \gamma^2\biggl]
	+ \biggl [\widetilde{\mu}_{+}^{(2)} \Bigl(3
	+ \frac{104}{7} \nu
	-  \frac{198}{7} \nu^2\Bigr)
	\nonumber\\
	& \qquad + \Delta \,\widetilde{\mu}_{-}^{(2)} \Bigl(3-  \frac{38}{7} \nu \Bigr)
	+ \frac{128}{3} \nu \widetilde{\sigma}_{+}^{(2)}\biggl] \gamma^6
	+ \biggl [\widetilde{\mu}_{+}^{(2)} \Bigl(- \frac{19}{2}
	+ \frac{617}{42} \nu
	+ \frac{5039}{42} \nu^2
	+ \frac{260}{21} \nu^3\Bigr)
	\nonumber\\
	& \qquad + \Delta \,\widetilde{\mu}_{-}^{(2)} \Bigl(- \frac{19}{2}
	+ \frac{1291}{42} \nu -  \frac{1649}{42} \nu^2\Bigr)
	+ \widetilde{\sigma}_{+}^{(2)} \Bigl(- \frac{64}{9} \nu
	-  \frac{1696}{7} \nu^2\Bigr)
	+ \frac{2048}{63} \Delta \nu \widetilde{\sigma}_{-}^{(2)}\biggl] \gamma^7\biggl\}\, ,\\[1ex]
	& I_{ijk}=m\nu r^3
	\Biggl[ n^{\langle i} n^{j} n^{k\rangle} \biggl\{- \Delta \Bigl(1
	-  \nu \gamma \Bigr)
	+ 18 \widetilde{\mu}_{-}^{(2)} \gamma^5
	+ \biggl [\Delta \widetilde{\mu}_{+}^{(2)} \Bigl(- \frac{3}{2}
	+ 48 \nu \Bigr)
	+ \widetilde{\mu}_{-}^{(2)} \Bigl(- \frac{39}{2}
	- 60 \nu \Bigr)
	- 84 \Delta \widetilde{\sigma}_{+}^{(2)}+ 84 \widetilde{\sigma}_{-}^{(2)}\biggl]
	\gamma^6\biggl\}\nonumber \\
	& \qquad +n^{\langle i} \lambda^{j} \lambda^{k\rangle}
	\biggl\{- \Delta \Bigl(1 - 2 \nu \Bigr) \gamma
	+ \biggl [\Delta \widetilde{\mu}_{+}^{(2)} \Bigl(-39
	+ 36 \nu \Bigr)
	+ \widetilde{\mu}_{-}^{(2)} \Bigl(39
	- 42 \nu \Bigr)
	- 72 \Delta \widetilde{\sigma}_{+}^{(2)}
	+ 72 \widetilde{\sigma}_{-}^{(2)}\biggl] \gamma^6\biggl\}\Biggl]
	\, ,\\[1ex]
	&I_{ijkl}=m\nu r^4 n^{\langle i} n^{j} n^{k} n^{l\rangle} \biggl [1
	- 3 \nu
	+ \Bigl(18 \widetilde{\mu}_{+}^{(2)}
	- 18 \Delta \widetilde{\mu}_{-}^{(2)}\Bigr) \gamma^5\biggl]\, , \\
	& J_{ij}={}\sqrt{G} \,(mr)^{3/2} \,\ell^{\langle i} n^{j\rangle}
	\biggl\{- \Delta \nu \biggl [1
	+ \Bigl(\frac{25}{28}
	+ \frac{3}{14} \nu \Bigr) \gamma \biggl]
	+ \Bigl(-9 \Delta \nu \widetilde{\mu}_{+}^{(2)}
	+ 9 \nu \widetilde{\mu}_{-}^{(2)}
	+ 12 \Delta \widetilde{\sigma}_{+}^{(2)}
	+ 12 \widetilde{\sigma}_{-}^{(2)}\Bigr) \gamma^5
	\\ & \qquad + \biggl [\Delta \widetilde{\mu}_{+}^{(2)} \Bigl(\frac{663}{28} \nu
	+ \frac{117}{7} \nu^2\Bigr)
	+ \widetilde{\mu}_{-}^{(2)} \Bigl(- \frac{177}{7} \nu
	+ \frac{477}{14} \nu^2\Bigr)
	+ \Delta \widetilde{\sigma}_{+}^{(2)} \Bigl(-10
	-  \frac{690}{7} \nu \Bigr)
	+ \widetilde{\sigma}_{-}^{(2)} \Bigl(-10
	-  \frac{346}{7} \nu \Bigr)\biggl] \gamma^6\biggl\}\, ,\nonumber \\[1ex]
	&J_{ijk}={} \sqrt{G} \,(mr)^{5/2} \dfrac{\nu}{m} \,\ell^{\langle i} n^{j} n^{k\rangle} \biggl\{1
	- 3 \nu
	+ \biggl [\widetilde{\mu}_{+}^{(2)} \Bigl(21
	- 27 \nu \Bigr)
	- 12 \Delta \widetilde{\mu}_{-}^{(2)}
	+ 64 \widetilde{\sigma}_{+}^{(2)}\biggl] \gamma^5\biggl\}\,.
\end{align}
\end{subequations}

\section{Energy flux for quasi-circular orbits}\label{sec:Flux}

\subsection{Mode decomposition of the instantaneous part of the flux}

Our main goal is to obtain the GW energy flux $\mathcal{F}\equiv(\dd E/\dd t)^\text{GW}$ at the NNL/2PN order. When non-linear tail effects in the wave propagation are ignored, the resulting ``instantaneous'' flux $\mathcal{F}_{\text{inst}}$ at the NNL/2PN order is a mere quadratic form of the $(\ell+1)$-th time derivatives $\mathcal{I}_L^{(\ell+1)}(t)$ of the source moments $\mathcal{I}_L\equiv\{I_L,J_L\}$, with general term $\propto \mathcal{I}^{(\ell+1)}_L \mathcal{I}^{(\ell+1)}_L $. Knowing those moments, the computation of $\mathcal{F}_{\text{inst}}$ is straightforward but can be performed in a particularly convenient way by decomposing the STF tensors $\mathcal{I}_L$ into some orthogonal STF basis associated with a natural triad of the problem. This decomposition  induces a related mode splitting of the flux, which is essential to its EoB treatment~\cite{DIN09}. We will thus define here the flux modes more precisely and list their tidal parts at the NNL order. 

[EDIT] When applying the time derivatives to compute $\mathcal{I}_L^{(\ell+1)}$, we may resort to the equations of motion obtained in Paper~I~\cite{HFB20a}. However, contrary to what was stated there, the expression for the acceleration displayed in Eq. (C1) of that article holds for a coordinate system that differs from the harmonic coordinates employed in the present work. Indeed, in one of the PN potentials parametrizing the Lagrangian (namely $\hat{X}$), the accelerations were replaced by their values on shell, which is equivalent to applying a gauge transformation to the particle positions. Correcting for this gauge mismatch, the relative acceleration in the same coordinate system as the source multipole moments \eqref{eq:momentsCM} reads
\begin{equation}\label{acccorr}
a^i = a^i_\text{Paper I} + \frac{6 G^3 m^2}{c^4 r^8}\Bigl(\mu_{+}^{(2)}+\Delta\,\mu_{-}^{(2)}\Bigr) \left[\left( 4\frac{G m}{r} +63(nv)^2 -7v^2 \right)n^i-14 (nv)v^i\right]\,.
\end{equation}
Performing the computations with the acceleration \eqref{acccorr}, \emph{i.e.}, replacing the accelerations by on-shell values given by \eqref{acccorr} when computing the time derivatives of the moments, we find that the radiated energy flux is modified.\footnote{This shift of the acceleration was not correcly taken into account in a previous version of the paper, resulting in erroneous expressions for the energy flux, the phase in the time domain and in the stationary phase approximation.}

As before we adopt the moving triad $(\boldsymbol{n},\boldsymbol{\lambda}, \boldsymbol{\ell})$, with $\boldsymbol{\ell}=\boldsymbol{n}\times \boldsymbol{v}/|\boldsymbol{n}\times \boldsymbol{v}|=\boldsymbol{n}\times \boldsymbol{\lambda}$ representing the unit vector pointing towards the Newtonian angular momentum or, alternatively, the value of the former triad at ascending node, say $(\boldsymbol{n}_0,\boldsymbol{\lambda}_0,\boldsymbol{\ell}_0)$. By definition of the orbital phase for planar orbits, we have $\boldsymbol{n}=\cos \phi\,  \boldsymbol{n}_0+\sin \phi\, \boldsymbol{\lambda}_0$. Posing  $\boldsymbol{\m}=(\boldsymbol{n}+\di \boldsymbol{\lambda})\sqrt{2}$ [or $\boldsymbol{\m}_0=(\boldsymbol{n}_0+\di \boldsymbol{\lambda}_0)\sqrt{2}$], it is often useful, in three dimensions, to introduce instead the associated complex triads $(\boldsymbol{\m},\overline{\boldsymbol{\m}},\boldsymbol{\ell})$ [or $(\boldsymbol{\m}_0,\overline{\boldsymbol{\m}}_0,\boldsymbol{\ell}_0)$], where the bar denotes the complex conjugation. Notice the simple relations $\boldsymbol{\m}=\de^{-\di\phi} \boldsymbol{\m}_0$ and $\boldsymbol{\ell}=\boldsymbol{\ell}_0$ for non-spinning (planar) binaries. Our orthogonal (un-normalized) STF basis will then be chosen to be $(\alpha_L^{\ell m})_{\ell,|m|\leqslant \ell}$, with $\alpha_L^{\ell m}=\m^{\langle M} \ell^{L-M \rangle}$ for $0\leqslant m\leqslant \ell$, and $\alpha_L^{\ell m}=(-1)^m \overline{\m}^{\langle |M|} \ell^{L-|M| \rangle}$ for $-\ell\leqslant m<0$. The basis $\alpha_{0L}^{\ell m}$ is defined in a similar way. One can prove the orthogonality condition:
\begin{align}\label{eq:ortho}
\alpha_{L}^{\ell m}\overline{\alpha}_{L}^{\ell m'} = \frac{(\ell-m)!(\ell+m)!}{2^m \ell!(2\ell-1)!!}\delta_{m,m'}\,.
\end{align}
Any source multipole moment $\mathcal{I}_L$ may now be written as
\begin{align}\label{eq:ML}
\mathcal{I}_L = \sum_{|m|\leqslant \ell} \mathcal{I}_{\ell m}  \alpha_{L}^{\ell m}=\sum_{|m|\leqslant \ell} \mathcal{I}_{\ell m}  \alpha_{0L}^{\ell m}\de^{-\di m \phi} \, .
\end{align}
For circular orbits, the coefficients $\mathcal{I}_{\ell m}$ only depend on the orbital frequency $\omega$. Because $\dd \phi/\dd t=\omega$, with $\dot{\omega}\equiv\dd \omega/\dd t=\mathcal{O}(1/c^5)$ vanishing up to the 2PN order, differentiation of the expression~\eqref{eq:ML} is equivalent to the replacement $\alpha_{L}^{\ell m}\to -\di m \omega \alpha_{L}^{\ell m}$ (which removes the mode $m=0$). In particular, the component $m$ of $\mathcal{I}^{(\ell+1)}_L$ is proportional to $\de^{-\di m \phi}$, as is the mode $h_{\ell m}$  associated with the latter multipole at linear order in the decomposition of $h=h_+-\di h_\times$ into spin-weighted spherical harmonics of weight -2 for planar systems (see, \textit{e.g.}, Ref.~\cite{BFIS08} for further explanations), which is a simple way to see that the two decompositions coincide, apart from normalization factors that must disappear from observable quantities.

The flux $\mathcal{F}_{\text{inst}}$ for circular orbits is thus made of a sum of terms
\begin{align} \label{eq:fluxterm}
\mathcal{I}^{(\ell+1)}_L \mathcal{I}^{(\ell+1)}_L = \sum_{m=1}^\ell \frac{(\ell-m)!(\ell+m)!}{2^{m-1} \ell!(2\ell-1)!!}(m \omega)^{2\ell +2} |\mathcal{I}_{\ell m}|^2  + \mathcal{O}(\dot{\omega})\, ,
\end{align}
where we have used the orthogonality formula~\eqref{eq:ortho}, the definition of $\alpha_L^{\ell\, -|m|}$, as well as the reality condition for $\mathcal{I}_L$, \textit{i.e.}, $\mathcal{I}_{\ell\, -m}=(-1)^m\overline{\mathcal{I}_{\ell m}}$. The precise $\ell$-dependent global factors are given in Eq.~\eqref{eq:flux} below, after the replacement $\mathcal{U}^{(1)}\rightarrow \mathcal{I}^{(\ell+1)}$ for the instantaneous part of the flux. We investigate the tail part of the flux (which depends on the past of the system) in \cref{sec:tailmode}. 

Finally, the part of the instantaneous flux proportional to $|\mathcal{I}_{\ell m}|^2 $ for $1\leqslant m\leqslant \ell$ will be denoted $\mathcal{F}^{\ell m}_\text{inst}$ henceforth. To present these modes to NNL order in the case of quasi-circular orbits, we employ the invariant dimensionless PN parameter $x=( \frac{G m \omega}{c^3})^{2/3}$. The non-zero tidal corrections in the various modes are given by~\cite{erratum,Patil2024}
\begin{subequations}\label{eq:Ftidalmode}
\begin{align}
F_{\text{inst}}^{22}={}& \widetilde{\mu}_{+}^{(2)} \Bigl(1
		+ 4 \nu \Bigr)
		+ \Delta\widetilde{\mu}_{-}^{(2)}
		+ \biggl [\widetilde{\mu}_{+}^{(2)} \Bigl(- \frac{22}{21}
		-  \frac{653}{42} \nu
		+ \frac{155}{21} \nu^2\Bigr)
		+ \Delta\widetilde{\mu}_{-}^{(2)} \Bigl(- \frac{22}{21}
		+ \frac{305}{42} \nu \Bigr)
		+ \frac{224}{9} \nu \widetilde{\sigma}_{+}^{(2)}\biggl] x\nonumber\\
		& + \biggl [\widetilde{\mu}_{+}^{(2)} \Bigl(\frac{167}{54}
		-  \frac{4579}{756} \nu
		-  \frac{125347}{1764} \nu^2
		+ \frac{5123}{1323} \nu^3\Bigr)
		+ \Delta\widetilde{\mu}_{-}^{(2)} \Bigl(\frac{167}{54}
		-  \frac{46937}{5292} \nu
		+ \frac{55985}{5292} \nu^2\Bigr)
		\nonumber\\
		& \quad + \widetilde{\sigma}_{+}^{(2)} \Bigl(- \frac{284}{63}\nu -  \frac{376}{189} \nu^2\Bigr)
		+ \frac{8084}{189} \Delta\widetilde{\sigma}_{-}^{(2)} \nu
		+ \frac{80}{3} \widetilde{\mu}_{+} ^{(3)} \nu \biggl] x^2\,,\\
F_{\text{inst}}^{21}={}&\biggl [\widetilde{\mu}_{+}^{(2)} \Bigl(\frac{1}{6} \nu
	-  \frac{2}{3} \nu^2\Bigr)
	-  \frac{1}{12} \Delta\widetilde{\mu}_{-}^{(2)} \nu
	+ \widetilde{\sigma}_{+}^{(2)} \Bigl(- \frac{1}{9}
	+ \frac{4}{9} \nu \Bigr)
	-  \frac{1}{9} \Delta\widetilde{\sigma}_{-}^{(2)}\biggl] x
	+ \biggl [\widetilde{\mu}_{+}^{(2)} \Bigl(\frac{5}{112} \nu
	-  \frac{5}{18} \nu^2
	+ \frac{25}{63} \nu^3\Bigr)\nonumber\\
	& \quad + \Delta\widetilde{\mu}_{-}^{(2)} \Bigl(\frac{1}{6} \nu
	-  \frac{35}{36} \nu^2\Bigr)
	+ \widetilde{\sigma}_{+}^{(2)} \Bigl(- \frac{173}{756}
	+ \frac{439}{189} \nu
	-  \frac{152}{27} \nu^2\Bigr)
	+ \Delta\widetilde{\sigma}_{-}^{(2)} \Bigl(- \frac{173}{756}
	+ \frac{32}{63} \nu \Bigr)\biggl] x^2\,,\\
F_{\text{inst}}^{33}={}&\biggl [\widetilde{\mu}_{+}^{(2)} \Bigl(\frac{3645}{448} \nu
	-  \frac{3645}{112} \nu^2\Bigr)
	-  \frac{3645}{448} \Delta\widetilde{\mu}_{-}^{(2)} \nu \biggl] x
	+ \biggl [\widetilde{\mu}_{+}^{(2)} \Bigl(- \frac{27945}{448} \nu
	+ \frac{244215}{896} \nu^2
	-  \frac{20655}{224} \nu^3\Bigr)\nonumber\\
	& \quad + \Delta\widetilde{\mu}_{-}^{(2)} \Bigl(\frac{13365}{224} \nu
	-  \frac{15795}{224} \nu^2\Bigr)
	+ \widetilde{\sigma}_{+}^{(2)} \Bigl(\frac{10935}{224} \nu
	-  \frac{10935}{56} \nu^2\Bigr)
	-  \frac{1215}{224} \Delta\widetilde{\sigma}_{-}^{(2)} \nu \biggl] x^2\,,\\
F_{\text{inst}}^{32}={}&\biggl [\widetilde{\mu}_{+}^{(2)} \Bigl(\frac{20}{21} \nu
	-  \frac{100}{21} \nu^2
	+ \frac{40}{7} \nu^3\Bigr)
	+ \Delta\widetilde{\mu}_{-}^{(2)} \Bigl(- \frac{20}{63} \nu
	+ \frac{20}{21} \nu^2\Bigr)
	+ \widetilde{\sigma}_{+}^{(2)} \Bigl(\frac{320}{189} \nu
	-  \frac{320}{63} \nu^2\Bigr)\biggl] x^2\,,\\
F_{\text{inst}}^{31}={}&\biggl [\widetilde{\mu}_{+}^{(2)} \Bigl(\frac{1}{1344} \nu
	-  \frac{1}{336} \nu^2\Bigr)
	-  \frac{1}{1344} \Delta\widetilde{\mu}_{-}^{(2)} \nu \biggl] x
	+ \biggl [\widetilde{\mu}_{+}^{(2)} \Bigl(- \frac{23}{12096} \nu
	+ \frac{107}{24192} \nu^2
	+ \frac{11}{864} \nu^3\Bigr)
	\nonumber\\
	& \quad + \Delta\widetilde{\mu}_{-}^{(2)} \Bigl(\frac{1}{432} \nu-  \frac{13}{6048} \nu^2\Bigr)
	+ \widetilde{\sigma}_{+}^{(2)} \Bigl(\frac{17}{2016} \nu
	-  \frac{17}{504} \nu^2\Bigr)
	-  \frac{1}{224} \Delta\widetilde{\sigma}_{-}^{(2)} \nu \biggl] x^2\,,\\
F_{\text{inst}}^{44}={}&\biggl [\widetilde{\mu}_{+}^{(2)} \Bigl(\frac{2560}{81} \nu
	-  \frac{28160}{189} \nu^2
	+ \frac{10240}{63} \nu^3\Bigr)
	+ \Delta\widetilde{\mu}_{-}^{(2)} \Bigl(- \frac{2560}{189} \nu
	+ \frac{2560}{63} \nu^2\Bigr)\biggl] x^2\,,\\
F_{\text{inst}}^{42}={}&\biggl [\widetilde{\mu}_{+}^{(2)} \Bigl(\frac{10}{567} \nu
	-  \frac{110}{1323} \nu^2
	+ \frac{40}{441} \nu^3\Bigr)
	+ \Delta\widetilde{\mu}_{-}^{(2)} \Bigl(- \frac{10}{1323} \nu
	+ \frac{10}{441} \nu^2\Bigr)\biggl] x^2\,.
\end{align}
\end{subequations}
For all modes we pose $\mathcal{F}_{\text{inst}}^{\ell m} = \frac{192c^5}{5 G} \nu\,x^{10} F_{\text{inst}}^{\ell m}$.

\subsection{Mode calculation of the tail part of the flux}\label{sec:tailmode}

The full flux, including non-linear propagation effects, 
is parametrized by the so-called mass and current radiative multipole moments $U_L(t)$ and $V_L(t)$ as
\begin{equation}\label{eq:flux}
\mathcal{F} = \dfrac{G}{c^{5}}\left\{ \dfrac{1}{5}U_{ij}^{(1)}U_{ij}^{(1)} + \dfrac{1}{c^{2}} \left[ \dfrac{1}{189} U_{ijk}^{(1)}U_{ijk}^{(1)} + \dfrac{16}{45}V_{ij}^{(1)}V_{ij}^{(1)}  \right] + \dfrac{1}{c^{4}} \left[ \dfrac{1}{9072}U_{ijkm}^{(1)}U_{ijkm}^{(1)} + \dfrac{1}{84}V_{ijk}^{(1)}V_{ijk}^{(1)} \right] + \mathcal{O}\left(\dfrac{1}{c^{6}} \right)  \right\}\,,
\end{equation}
where we have restricted ourselves to the 2PN order. The physical content of this expression lies in the relationship between the radiative moments and the source moments computed in \cref{sec:MultipolarMoments}. At the linear level, the radiative moments $\mathcal{U}_L=\{U_L,V_L\}$ are just the $\ell$-th time derivatives of the source moments $\mathcal{I}_L=\{I_L,J_L\}$. At the quadratic level, the radiative moments involve the interaction between the ADM mass $M$ and the source moments $\mathcal{I}_L$ in the form of the non-local tail integrals~\cite{BD92}:\footnote{For the present calculation, we do not need to consider other non-tail (instantaneous) terms arising at the same order $\mathcal{O}(1/c^{3})$ as the tails but only for $\ell\geqslant 4$ in the mass sector and $\ell\geqslant 3$ in the current sector (see Ref.~\cite{BFIS08}).} 
\begin{subequations}\label{eq:radtail}
	\begin{align}
	U_L(t) &= I^{(\ell)}_L(t) +
	\frac{2G M}{c^3} \int^{+\infty}_0 \dd\tau \,
	I^{(\ell+2)}_L (t-\tau) \ln \left(
	\frac{\tau}{\tau_\ell} \right) +
	\mathcal{O}\left( \frac{1}{c^5} \right)\,,
	\\ V_L(t) &= J^{(\ell)}_L(t) +
	\frac{2G M}{c^3} \int^{+\infty}_0 \dd\tau \,
	J^{(\ell+2)}_L (t-\tau) \ln \left(
	\frac{\tau}{\lambda_\ell} \right) +
	\mathcal{O}\left( \frac{1}{c^5} \right)\,,
	\end{align}
\end{subequations}
where $\tau_\ell$ and $\lambda_\ell$ denote two gauge constants, which will cancel out in the end of our calculation. Consistently with the approximation, we include the leading and NL tail effects which will correspond to formal 6.5PN and 7.5PN contributions in the tidal terms. We then need only the tail entering the mass quadrupole, current quadrupole and mass octupole moments.\footnote{The expression of the link between the radiative moments $\mathcal{U}_L$ and the source moments $\mathcal{I}_L$ is known to simplify significantly when appropriate mass and current canonical moments, $M_L=I_L+\mathcal{O}(1/c^5)$ and $S_L=J_L+\mathcal{O}(1/c^5)$, are introduced instead of the source moments (see Eq. (5.9) of Ref.~\cite{BFIS08}). We checked that the leading correction in the canonical mass quadrupole moment $M_{ij}$ for which tidal effects would give a contribution at the NNL/7.5PN level actually vanishes. On the other hand, the radiation reaction dissipative pieces $\mathcal{O}(\dot{\omega})$ being purely instantaneous and ``time-odd'', cannot contribute to the flux for quasi-circular orbits at this level.}

The computation of the tails is conveniently achieved by starting from the following alternative form for the instantaneous and tail part of the radiative moments: 
\begin{align} \label{eq:insttail}
\mathcal{U}^{\text{inst+tail}}_L(t) = \mathcal{I}_L^{(\ell)}(t) +   \frac{2 G M}{c^3} \Bigg[ \ln\bigg(\frac{\mathcal{T}}{P_\ell}\bigg) \,\mathcal{I}^{(\ell+1)}(t) + \int_0^\mathcal{T} \dd \tau\, \ln \tau \,\mathcal{I}_L^{(\ell+2)}(t-\tau) + \int_\mathcal{T}^{+\infty}\frac{\dd\tau}{\tau} \,\mathcal{I}_L^{(\ell+1)}(t-\tau)\Bigg]\, ,
\end{align}
where $\mathcal{T}$ is an arbitrary time scale and $P_\ell$ denotes either $\tau_\ell$ or $\lambda_\ell$. Now, it was proved in the Appendix B of~\cite{BS93} that, in the case of decaying quasi-circular orbits, the frequency on which depend the integrands in \cref{eq:insttail}, \textit{e.g.} $(\dd/\dd t)^{(\ell+1)}[\mathcal{I}_{\ell m}(t-\tau)\alpha_{0L}^{\ell m}\de^{- \di m \phi(t-\tau)}]$, can be substituted with its value at the current time $t$, modulo some remainder $\mathcal{O}(\ln c/c^5)$. This amounts to replacing the frequency $\omega(t-\tau)$ by $\omega(t)$ and the phase $\phi(t-\tau)$ by $\phi(t)-\omega(t)\tau$. The expression inside the square brackets in Eq.~\eqref{eq:insttail} then reads
\begin{align}
\sum_{0<|m|\leqslant \ell}(-\di m \omega)^{\ell+1}\mathcal{I}_{\ell m}\bigg[\ln\bigg(\frac{\mathcal{T}}{P_\ell}\bigg)  + (-\di m \omega)\int_0^\mathcal{T} \dd \tau\, \ln \tau\,  \de^{\di m \omega \tau}+ \int_{\mathcal{T}}^{+\infty}\frac{\dd\tau}{\tau} \de^{\di m \omega \tau} \bigg] \alpha_L^{\ell m}+\mathcal{O}\left(\frac{\ln c}{c^5}\right)\,.  
\end{align}
After explicit integration we find
\begin{align} \label{eq:insttailexplicit}
\mathcal{U}^{\text{inst+tail}}_L = \sum_{0<|m|\leqslant \ell}(-\di m \omega)^{\ell}\mathcal{I}_{\ell m}\bigg[1- \frac{2 G M \di m \omega}{c^3} c_{\ell m}(\omega)\bigg] + \mathcal{O}\left(\frac{\ln c}{c^5}\right)\, ,
\end{align}
with $c_{\ell m}(\omega)= \di \text{ sign}(m) \,\pi/2 -(\ln (|m|\omega P_\ell)+\gamma_\text{E}) $. The flux is obtained by squaring Eq.~\eqref{eq:insttailexplicit} multiplied by an extra factor $-\di m\omega$, keeping only the leading $M \times \mathcal{I}_L$ correction. Exploiting the orthogonality relation~\eqref{eq:ortho} yields 
\begin{align}
\mathcal{F}_{\text{inst+tail}} &\propto \sum_{m=1}^\ell \frac{(\ell-m)!(\ell+m)!}{2^{m-1} \ell!(2\ell-1)!!}(m \omega)^{2\ell +2} \bigg[1+\frac{2G M m \omega}{c^3}\di (c_{\ell m}(\omega)- \overline{c}_{\ell m}(\omega))\bigg] + \mathcal{O}\left(\frac{1}{c^6}\right)
\nonumber \\ &  = \sum_{m=1}^\ell \mathcal{F}_{\text{inst}}^{\ell m} \bigg[1+\frac{2\pi G M\omega}{c^3} m\bigg]+ \mathcal{O}\left(\frac{1}{c^6}\right)\, .
\end{align}
Thus, the tail contribution of the $(\ell, m)$ flux piece at the relative 1.5PN order is simply given by $2\pi k_m \mathcal{F}_{\text{inst}}^{\ell m}$ with $k_m = G \omega m M/c^3$. Note that the factor $k_m$ is just the first order term in the expansion of the squared module of the tail resummed-factor $T_{\ell m}$ introduced in Ref.~\cite{DIN09}. The ADM mass $M$ must crucially include the leading tidal corrections. For the tidal tail at the NNL/7.5PN order, it is sufficient to take (with here $m=m_1+m_2$)
\begin{align}\label{eq:Madm}
M = m + \frac{E}{c^2} = m \bigg[1 -\frac{\nu x}{2}\Bigl(1-18\,\widetilde{\mu}^{(2)}_+ x^5\Bigr)\bigg] + \mathcal{O}\left(\frac{1}{c^4}\right)\, .
\end{align}
To end up, let us provide the resulting tail part of the flux for circular orbits at the NNL/7.5PN order for tidal effects:
\begin{align}\label{eq:Ftail}
\mathcal{F}_\text{tail} =& \dfrac{768\pi c^5}{5G} \nu \,x^{23/2} \Biggl\{   (1+4 \nu)\widetilde{\mu}_{+}^{(2)} + \Delta \widetilde{\mu}_{-}^{(2)}  \nonumber\\
& ~ +  \left[\left(-\dfrac{22}{21} - \dfrac{5053}{1344} \nu - \dfrac{2029}{48}\nu^2 \right)\widetilde{\mu}_{+}^{(2)} + \Delta \left( - \dfrac{22}{21} - \dfrac{351}{64} \nu \right) \widetilde{\mu}_{-}^{(2)} + \left(-\dfrac{1}{18} + \dfrac{226}{9} \nu  \right)\widetilde{\sigma}_{+}^{(2)} -\dfrac{\Delta}{18} \widetilde{\sigma}_{-}^{(2)}  \right]x\Biggr\}\,.
\end{align}

\section{GW phase evolution for quasi-circular orbits}\label{sec:Phase}

Writing the GW energy flux as $\mathcal{F} = \mathcal{F}_{\text{pp}} + \mathcal{F}_{\text{tidal}}$, we have just computed the tidal part of the dissipative energy flux, $\mathcal{F}_{\text{tidal}}$,  in Eqs.~\eqref{eq:Ftidalmode} and \eqref{eq:Ftail}. The part generated by point particles without internal structure, $\mathcal{F}_{\text{pp}}$, already known~\cite{BDIWW95,BDI95,WW96,B96}, is given to consistent order by 
\begin{equation}\label{eq:Fpp}
\mathcal{F}_{\text{pp}} = \dfrac{32 c^5 \nu^2 x^5}{5G} \left\{ 1 + \left( -\dfrac{1247}{336}-\dfrac{35}{12} \nu \right)x +4\pi x^{3/2} + \left( -\dfrac{44711}{9072} + \dfrac{9271}{504}\nu +\dfrac{65}{18} \nu^2 \right)x^{2} + \left( -\dfrac{8191}{672} - \dfrac{583}{24}\nu \right) \pi x^{5/2} \right\}\,.
\end{equation}
Then, the tidal contribution to the flux complete to NNL order (including leading and NL tail terms) reads~\cite{erratum}\footnote{We recall that the polarizability coefficients are defined by Eqs.~\eqref{eq:polarpm}--\eqref{eq:polarpmtilde}. Note that the prefactors in front of~\eqref{eq:Fpp} and~\eqref{eq:Ftidal} are different.}
\begin{align}\label{eq:Ftidal}
\mathcal{F}_{\text{tidal}} &= \dfrac{192 c^5 \nu\,x^{10}}{5G} \Biggl\{ (1+4 \nu)\widetilde{\mu}_{+}^{(2)} + \Delta \widetilde{\mu}_{-}^{(2)} \\
& + \left[ \left(-\dfrac{22}{21} -\dfrac{1217}{168} \nu - \dfrac{155}{6}\nu^2 \right)\widetilde{\mu}_{+}^{(2)} + \Delta \, \left( -\dfrac{22}{21} - \dfrac{23}{24}\nu \right)\widetilde{\mu}_{-}^{(2)} + \left(-\dfrac{1}{9} +\dfrac{76}{3}\nu \right)\widetilde{\sigma}_{+}^{(2)} -\dfrac{1}{9}\Delta \widetilde{\sigma}_{-}^{(2)}  \right] x \nonumber\\
& +4\pi \left[ (1+4 \nu)\widetilde{\mu}_{+}^{(2)} + \Delta \widetilde{\mu}_{-}^{(2)} \right] x^{3/2} \nonumber \\
& + \left[\left(\dfrac{167}{54} - \dfrac{649853}{18144} \nu + \dfrac{15923}{336}\nu^2 +\dfrac{965}{12} \nu^3 \right)\widetilde{\mu}_{+}^{(2)} + \Delta \left( \dfrac{167}{54} + \dfrac{74783}{2016} \nu - \dfrac{2779}{144} \nu^2 \right) \widetilde{\mu}_{-}^{(2)} \right. \nonumber\\
& \qquad\left. + \left(-\dfrac{173}{756} + \dfrac{145}{3} \nu -208\nu^2 \right)\widetilde{\sigma}_{+}^{(2)} + \Delta \left(-\dfrac{173}{756} + \dfrac{1022}{27} \nu \right) \widetilde{\sigma}_{-}^{(2)} + \dfrac{80}{3}\nu \widetilde{\mu}_{+}^{(3)} \right]x^2 \nonumber \\
&+ 4\pi \left[\left(-\dfrac{22}{21} - \dfrac{5053}{1344} \nu - \dfrac{2029}{48}\nu^2 \right)\widetilde{\mu}_{+}^{(2)} + \Delta \left( - \dfrac{22}{21} - \dfrac{351}{64} \nu \right) \widetilde{\mu}_{-}^{(2)} 
+ \left(-\dfrac{1}{18} + \dfrac{226}{9} \nu  \right)\widetilde{\sigma}_{+}^{(2)} -\dfrac{\Delta}{18} \widetilde{\sigma}_{-}^{(2)}  \right]x^{5/2}\Biggr\}\,.\nonumber
\end{align}
We find agreement on the terms proportional to $\widetilde{\mu}_{\pm}^{(2)}$ with the new results from EFT method~\cite{Patil2024}.

Together with the conservative energy of the system available in Eqs.~(6.5) of~\cite{HFB20a}, the above energy flux permits determining the frequency and phase evolution for circular orbits through the two ordinary differential equations
\begin{equation}
\frac{\dd\omega}{\dd t} = - \frac{\mathcal{F}(\omega)}{\dd E/\dd\omega}\,,\qquad
\frac{\dd\varphi}{\dd t} = \omega\,.
\end{equation}
As is well known, there are various ways to solve those equations approximately, called PN approximants, yielding significant deviations from numerical relativity at small separations, \textit{i.e.}, outside the domain of validity of the PN expansion~\cite{BuonSathya15}. Following the simplest adiabatic Taylor PN approximant, we obtain the phase in the time domain as $\varphi = \varphi_{\text{pp}} + \varphi_{\text{tidal}}$, where we recall the point-particle result up to 2.5PN order,
\begin{align}
\varphi_{\text{pp}} &= - \dfrac{1}{32 \nu x^{5/2}} \biggl\{ 1 + \left( \dfrac{3715}{1008}+\dfrac{55}{12} \nu \right)x- 10\pi x^{3/2} + \left( \dfrac{15293365}{1016064} + \dfrac{27145}{1008}\nu +\dfrac{3085}{144} \nu^2 \right)x^{2} \nonumber\\&\qquad\qquad\qquad + \left( \dfrac{38645}{1344} - \dfrac{65}{16}\nu \right) \pi x^{5/2} \ln\left(\frac{x}{x_0}\right)\biggr\} \,,
\end{align}
and where the tidal contribution is~\cite{erratum}
\begin{align}
\varphi_{\text{tidal}} =& - \dfrac{3 x^{5/2}}{16 \nu^2} \Biggl\{ (1+22 \nu)\widetilde{\mu}_{+}^{(2)} + \Delta \widetilde{\mu}_{-}^{(2)}  \\
& + \left[ \left(\dfrac{195}{56} + \dfrac{1595}{14} \nu + \dfrac{325}{42}\nu^2 \right)\widetilde{\mu}_{+}^{(2)} + \Delta \, \left( \dfrac{195}{56} + \dfrac{4415}{168}\nu \right)\widetilde{\mu}_{-}^{(2)} + \left(-\dfrac{5}{63} +\dfrac{3460}{21}\nu \right)\widetilde{\sigma}_{+}^{(2)}  -\dfrac{5}{63} \Delta \widetilde{\sigma}_{-}^{(2)}  \right] x \nonumber\\
& -\dfrac{5\pi}{2} \left[(1+22 \nu)\widetilde{\mu}_{+}^{(2)} + \Delta \widetilde{\mu}_{-}^{(2)} \right] x^{3/2} \nonumber\\
& + \left[\left(\dfrac{136190135}{9144576} + \dfrac{978554825}{1524096} \nu - \dfrac{281935}{2016}\nu^2 +5 \nu^3 \right)\widetilde{\mu}_{+}^{(2)} + \Delta \left( \dfrac{136190135}{9144576} + \dfrac{213905}{864} \nu + \dfrac{1585}{432} \nu^2 \right) \widetilde{\mu}_{-}^{(2)}\right. \nonumber\\
& \qquad \left. + \left(-\dfrac{745}{1512} + \dfrac{1933490}{1701} \nu - \dfrac{3770}{27}\nu^2  \right)\widetilde{\sigma}_{+}^{(2)} + \Delta \left(-\dfrac{745}{1512} + \dfrac{19355}{81} \nu \right)\widetilde{\sigma}_{-}^{(2)} + \dfrac{1000}{9}\nu \widetilde{\mu}_{+}^{(3)} \right]x^2  \nonumber\\
& + \pi \left[ \left(-\dfrac{397}{32} - \dfrac{5343}{16} \nu + \dfrac{1315}{12}\nu^2 \right)\widetilde{\mu}_{+}^{(2)} + \Delta \, \left( -\dfrac{397}{32} - \dfrac{6721}{96}\nu \right)\widetilde{\mu}_{-}^{(2)} + \left(\dfrac{1}{3} -\dfrac{4156}{9}\nu \right)\widetilde{\sigma}_{+}^{(2)}  + \dfrac{\Delta}{3} \widetilde{\sigma}_{-}^{(2)} \right]x^{5/2}\Biggr\}\,.\nonumber
\end{align}

Next, motivated by data analysis applications, we provide the phase in the Fourier domain within the stationary-phase approximation for the dominant mode at twice the orbital frequency, with Fourier GW frequency $f$ and PN parameter $v \equiv (\frac{\pi G m f}{c^3})^{1/3}$. We find for this phase: $ \psi^\text{SPA} = 2 \pi f \,t_\text{c} + \psi_\text{pp} + \psi_\text{tidal}$ with~\cite{erratum}
\begin{subequations}\label{eq:SPA}
\begin{align}
\psi_{\text{tidal}} =& -\dfrac{9 v^{5}}{16 \nu^2} \Biggl\{ (1+22 \nu)\widetilde{\mu}_{+}^{(2)} + \Delta \widetilde{\mu}_{-}^{(2)} \\
& + \left[ \left(\dfrac{195}{112} + \dfrac{1595}{28} \nu + \dfrac{325}{84}\nu^2 \right)\widetilde{\mu}_{+}^{(2)} + \Delta \, \left( \dfrac{195}{112} + \dfrac{4415}{336}\nu \right)\widetilde{\mu}_{-}^{(2)} + \left(-\dfrac{5}{126} +\dfrac{1730}{21}\nu \right)\widetilde{\sigma}_{+}^{(2)}  -\dfrac{5}{126} \Delta \widetilde{\sigma}_{-}^{(2)}  \right] v^{2} \nonumber\\
& -\pi \left[ (1+22 \nu)\widetilde{\mu}_{+}^{(2)} + \Delta \widetilde{\mu}_{-}^{(2)} \right]v^{3} \nonumber\\
& + \left[\left(\dfrac{136190135}{27433728} + \dfrac{978554825}{4572288} \nu - \dfrac{281935}{6048}\nu^2 + \dfrac{5}{3} \nu^3 \right)\widetilde{\mu}_{+}^{(2)} + \Delta \left( \dfrac{136190135}{27433728} + \dfrac{213905}{2592} \nu + \dfrac{1585}{1296} \nu^2 \right) \widetilde{\mu}_{-}^{(2)}\right. \nonumber\\
& \qquad \left. + \left(-\dfrac{745}{4536} + \dfrac{1933490}{5103} \nu - \dfrac{3770}{81}\nu^2  \right)\widetilde{\sigma}_{+}^{(2)} + \Delta \left(-\dfrac{745}{4536} + \dfrac{19355}{243} \nu \right)\widetilde{\sigma}_{-}^{(2)} + \dfrac{1000}{27}\nu \widetilde{\mu}_{+}^{(3)} \right]v^{4} \nonumber\\
& + \pi \left[ \left(-\dfrac{397}{112} - \dfrac{5343}{56} \nu + \dfrac{1315}{42}\nu^2 \right)\widetilde{\mu}_{+}^{(2)} + \Delta \, \left( -\dfrac{397}{112} - \dfrac{6721}{336}\nu \right)\widetilde{\mu}_{-}^{(2)} + \left(\dfrac{2}{21} -\dfrac{8312}{63}\nu \right)\widetilde{\sigma}_{+}^{(2)}  + \dfrac{2}{21}\Delta \widetilde{\sigma}_{-}^{(2)} \right]v^{5}\Biggr\} \nonumber\,.
\end{align}
\end{subequations}
The result for the tidal part of the SPA phase in the case of equal bodies, with the same mass and identical polarizability parameters, has already been provided in Eq.~\eqref{eq:phiequal}.

\section{Summary and conclusions}\label{sec:conclusion}

In this paper and the preceding one~\cite{HFB20a}, we have solved the problem of the dynamics and GW emission of compact binary systems without spins for tidal, internal structure-dependent effects at the next-to-next-to-leading (NNL) order, meaning formally the order 7.5PN (taking into account tails) in the GW phase evolution. We used the formalism of the effective matter action of Ref.~\cite{BiniDF12}, which describes massive point-like particles with internal structure by introducing specific non-minimal couplings to the space-time curvature that model the finite size effects of the compact bodies due to the tidal interactions. Since the matter action is localized on the worldline of the particles, it is sometimes referred to as a ``skeletonized'' action. To the NNL order there appear three polarizability coefficients corresponding to mass quadrupole, current quadrupole and mass octupole tidal interactions. In Ref.~\cite{HFB20a}, we derived the associated effective Fokker action to obtain the conservative dynamics, \textit{i.e.}, EoM and conserved integrals of the motion.

In the present paper, we computed the matter stress-energy tensor of the compact binary from the same effective action, and inserted it into a GW generation formalism based on MPM approximations for the external field~\cite{BD86}, which are matched to the PN expansion of the inner field~\cite{B98mult, PB02}. The MPM-PN approach constitutes a very general way for computing the GW emission (and radiation reaction onto the source) once one is given the matter stress-energy-tensor. In particular, we resorted to general ready-to-use expressions for the source multipole moments and nonlinear interactions between those moments (tails, \textit{etc.}) leading to the observable waveform at infinity and, thus, the energy flux. At last, once the flux to NNL order for tidal effects had been obtained and reduced for circular orbits, we combined it with the result for the conservative energy found in~\cite{HFB20a}. Namely, we employed the standard flux-balance argument to determine the binary's chirp, \textit{i.e.}, the orbital phase and frequency evolution through GW emission for compact binaries on quasi-circular orbits. 

Our results extend and complete several previous results in the literature. In the Table~\ref{tab:literature}, we summarize the previous achievements in the field for each PN order and multipole component. We agree with all the previous results quoted in Table~\ref{tab:literature}. Finally, with the present paper, the tidal phase of non-spinning NS binaries is complete up to the NNL order including NL tails, which means formally up to the high 7.5PN level.\footnote{However we disagree with some coefficients in the literature. First, with the 6PN coefficient due to the current quadrupole moment computed in Ref.~\cite{Landry18}. Second, with the mass quadrupole contribution to the tail term at the 7.5PN order as reported in Ref.~\cite{DNV12}. The latter reference obtains for the mass quadrupole contributions to the SPA phase of two identical NS (see Eq.~(31) in~\cite{DNV12}):
$$\psi_\text{tidal}^\text{DNV} = - \kappa_{2}^{T}\frac{39}{4}v^{5} \left[ 1+\frac{3115}{1248}v^2 - \pi v^{3} + \left( \frac{23073805}{3302208} + \frac{20}{81} \bar{\alpha}^{(2)}_{2} + \frac{20}{351} \beta^{22}_{2} \right)v^{4} - \frac{4283}{1092}\pi v^{5} \right]\,,$$
with $\kappa_{2}^{T} = 6 \widetilde{\mu}^{(2)}_{+}$ in their notation (recall that we have $\widetilde{\mu}^{(2)}_{-}=0$ for identical NS). Further work~\cite{BiniDF12} fixed $\bar{\alpha}^{(2)}_{2}=85/14$ to be the contribution of the NNL equations of motion to the phasing. Now, the comparison with our present results, given for two equal bodies by Eq.~\eqref{eq:phaseSPA} in the Introduction, permits inferring that $\beta^{22}_{2}= 1150115/1016064$, so that, with this value, we are in agreement up to the NNL level for the mass quadrupole interaction; but we find that the NL 7.5PN tail term has the coefficient $-\frac{2137}{546}\pi \simeq -12.296$ instead of the coefficient $-\frac{4283}{1092}\pi \simeq -12.322$ obtained in Ref.~\cite{DNV12}.}
\begin{table}[htb]
	\begin{tabular}{|c||c|c|c|}
\hline $\varphi_\text{tidal}$ & Mass quadrupole & Current quadrupole & Mass octupole \\ 
\hline \hline 5PN (L) & ~\cite{FH08, VHF11, DNV12, VF13, F14} \checkmark & ~$\times$ & ~$\times$\\ 
\hline 6PN (NL) & ~\cite{DNV12, VF13, AGP18} \checkmark & ~\cite{AGP18, BV18} \checkmark & ~$\times$\\ 
\hline 6.5PN (tail) & ~\cite{DNV12, AGP18} \checkmark & ~$\times$ & ~$\times$\\ \hline  
\hline 7PN (NNL) & \checkmark & \checkmark & \cite{AGP18,Landry18} \checkmark \\ \hline  
\hline 7.5PN (tail) & \checkmark & \checkmark & ~$\times$ \\ \hline  
\end{tabular}
\caption{Comparison with the existing literature. We indicate for each order and each multipolar piece contributing to the tidal phase $\varphi_\text{tidal}$ the previous references having achieved it and with which we agree (note that Ref.~\cite{DNV12} considers only the case of equal bodies). The contributions obtained with the present paper are indicated as a checkmark \checkmark. Up to NNL order including tails, the tidal phase is now complete.}\label{tab:literature}
\end{table}
	
\acknowledgments

We thank Justin Vines for useful discussions. We thank Raj Patil for anticipating the results of~\cite{Patil2024}, computed via EFT Feynman diagrammatic methods, indicating an error in the flux, which we correct in the new version of this paper.

\appendix

\section{Proof that dimensional and Hadamard regularizations are equivalent at NNL order}\label{app:proof}

In order to show that dimensional and Hadamard regularizations are equivalent at NNL order, we need to show that the multipole moments integrated with these two regularizations have the same value. The regularizations appear in two different steps of the computation: the integration of the elementary potentials and the integration of the multipole moments. We deal in \cref{app:proofpotentials} with the required potentials and in \cref{app:proofmultipoles} with the multipoles. In order to do so, we give general arguments regarding the structure of the different quantities considered. In both parts of this proof, we use the fact that in $d=3+\varepsilon$ dimensions, the sources $\sigma$, $\sigma_i$ and $\sigma_{ij}$ displayed in \cref{eq:pppart,eq:sigmatidal} have the same structure as in 3d and that their value are continuous for $\varepsilon \rightarrow 0$.

\subsection{Equivalence for potentials}\label{app:proofpotentials}

As said in \cref{sec:potentials}, we only require the potentials $V$ at 1PN and $V_i$, $\hat{W}_{ij}$ at Newtonian order for the integration of the multipole moments. Their definitions are given in \cref{eq:potentials}. Both $V$ and $V_i$ have sources with compact supports and are thus qualified of ``compact support potentials'' (C). By contrast, the source of $\hat{W}_{ij}$ involves a compact part as well as a non-compact part, hence $\hat{W}_{ij}$ itself splits correspondingly into a compact potential and a so-called ``non-compact support potential'' (NC). The latter is actually a ``$\partial V\partial V$ potential'', defined in general as a NC potential whose source can be written as $S^{(\partial V \partial V)} =\partial^{p_A} A^{\text{(C)}} \partial^{p_B} B^{\text{(C)}}$, where  $A^{\text{(C)}}$ and  $B^{\text{(C)}}$ are two C potentials, while $p_A$ is the number of spatial or time derivatives considered. Both C or $\partial V\partial V$ types have to be treated differently since the structure of their terms are different. The differences between regularizations arise when we take the limit $r_1=\vert\mathbf{x}-\boldsymbol{y}_1\vert\to 0$. It is then very convenient to split for a function $F$ its regular part from its singular part when $r_1\rightarrow 0$. For any function $F$ admitting a power-like expansion (with possible powers of logarithms) when $r_1\rightarrow 0$, we can define its regular part in the neighborhood of 1 as that part of this expansion that is smooth (with $k\in\mathbb{N}$), \textit{i.e.},
\begin{align}\label{eq:Freg}
F_\text{reg}= \sum_{\ell \geqslant 0, k\geqslant 0} \mathop{\hat{f}}_1\!{}^{L}_{k} \hat{r}_1^L r_1^{2k},
\end{align}
where ${}_1 \hat{f}^{L}_{k}$ are STF tensor coefficients contracted with $\hat{r}_1^L=\text{STF} (r_1^{i_1} r_1^{i_2} ... r_1^{i_{\ell}})$. What remains in the expansion defines the so-called singular part $F_\text{sing}=F-F_\text{reg}$. The extraction of the regular part of $F$ at 1 defines an operator $R_1$ as $R_1 [F]=F_\text{reg}$. The properties of this operator and its counterpart $S_1$, such that $S_1[F]=F_\text{sing}$, are discussed in Sec.~IIIB of Ref.~\cite{BiniDF12}. Let us simply remind that they are linear and commute with space or time derivatives.

For any potential $P$ considered in our problem, having a C part $P^{\text{(C)}}$ and a ``$\partial V \partial V$'' part $P^{(\partial V\partial V)}$, it can be proved, by means of the techniques described in Refs.~\cite{BFP98, MHLMFB20}, that their regular and singular parts are of the form
\begin{subequations}
  \begin{align}
P^{\text{(C)}}_\text{sing}   &=  
 \sum_{0\leqslant k}\sum_{0 \leqslant j \leqslant 2k+n} \propto \frac{\partial_{J}r_1^{2k-1-\varepsilon}}{c^{2k}}\, , \label{eq:PCsing} \\ P^{\text{(C)}}_\text{reg}   &=  
 \sum_{0\leqslant k}\sum_{0 \leqslant j \leqslant 2k+n} \propto \frac{\partial_{J}r_2^{2k-1-\varepsilon}}{c^{2k}}\, , \label{eq:PCreg}\\
P^{(\partial V\partial V)}_{\text{sing}} &=  \mathop{\sum_{1\leqslant s_A\leqslant p_A+n_A}}_{1\leqslant s_B\leqslant p_B + n_B}  \Bigg\{ \mathop{\sum_{0\leqslant j\leqslant s_A+s_B}}_{j+s_A+s_B\text{ even}} \!\!\!\!\propto \frac{\hat{n}_1^J}{r_1^{s_A+s_B+2\varepsilon}}+ \propto \Big(\partial_{1S_A}\partial_{2S_B} g^{(3+\varepsilon)}_\text{sing}+ 1\leftrightarrow 2 \Big) \Bigg\}\, , \label{eq:PdVdVsing} \\
P^{(\partial V\partial V)}_{\text{reg}} &=  \mathop{\sum_{1\leqslant s_A\leqslant p_A+n_A}}_{1\leqslant s_B\leqslant p_B + n_B}  \Bigg\{ \mathop{\sum_{0\leqslant j\leqslant s_A+s_B}}_{j+s_A+s_B\text{ even}} \!\!\!\!\propto \frac{\hat{n}_2^J}{r_2^{s_A+s_B+2\varepsilon}} + \propto \Big(\partial_{1S_A}\partial_{2S_B} g^{(3+\varepsilon)}_\text{reg}+ 1\leftrightarrow 2 \Big) \Bigg\}\,, \label{eq:PdVdVreg}
\end{align}
\end{subequations}
where $p_A$ comes from the definition of $S^{(\partial V \partial V)}$, $n_A$ is the number of spatial derivatives in the source of $A^{\text{(C)}}$, and $\partial_{Ai} \equiv \partial/\partial y_A^i$. The functions $g^{(3+\varepsilon)}_\text{reg}$ and $g^{(3+\varepsilon)}_\text{sing}$ are given by
\begin{subequations}
\begin{align}\label{eq:gsing}
g^{(d)}_\text{sing} &= S_1[g^{(d)}]= r_{12}^{-2\varepsilon} \sum_{\ell=0}^{+\infty}\frac{c^{(3+\varepsilon)}_{\ell}}{\ell+1} \hat{n}_{12}^L \hat{n}_{1}^L \left(\frac{r_1}{r_{12}}\right)^{\ell+1-\varepsilon}\,,\\ g^{(d)}_\text{reg} &= R_1[g^{(d)}]= -r_{12}^{-2\varepsilon} \sum_{\ell=0}^{+\infty}\frac{c^{(3+\varepsilon)}_{\ell}}{\ell+\varepsilon} \left(\frac{\hat{n}_{12}^L}{r_{12}^\ell}\right) \hat{r}_{1}^L +\frac{1}{2(1+\varepsilon)} \frac{r_0^{-2\varepsilon}}{\varepsilon}\,,\label{eq:greg}
\end{align}
\end{subequations}
with $c^{(d)}_{\ell} =(-1)^\ell 2^{\ell-1} \Gamma(d/2+\ell-1)/[\ell!\, \Gamma(d/2-1)(1-\varepsilon)]$. In order to show that the potentials have the same expression for both regularizations, we have to check that the structures of the regular and singular part are the same for $\varepsilon=0$ and $\varepsilon\neq 0$, meaning that poles in $1/\varepsilon$ do not appear.

\subsubsection{Compact support}

$P^{\text{(C)}}$ evaluated at $\mathbf{x}=\boldsymbol{y}_1$ in dimensional regularization or directly in 3 dimensions with the help of Hadamard's procedure lead to the same result, for in neither cases the purely singular part contributes. Indeed, in dimensional regularization, for an appropriate choice of $\varepsilon$, it automatically vanishes for $r_1=0$. In the Hadamard's finite part regularization, the reason for which it is zero is instead that the finite part of a function $F$ does belong to $R_1[F]$.

\subsubsection{Non-compact support}

Let us now demonstrate that the two terms in \cref{eq:PdVdVsing} are equivalent for both regularizations. The potential $P^{(\partial V \partial V)}$ is sourced by
\begin{equation}\label{eq:SdVdV}
S^{(\partial V \partial V)}  =\mathop{\sum_{s_A\leqslant p_A+n_A}}_{s_B\leqslant p_B + n_B} \Bigg\{ \propto  \frac{n_1^{S_A S_B}}{r_1^{2 +s_A+s_B+ 2\varepsilon}} + \left(\propto \partial_{S_A}\bigg(\frac{1}{r_1^{1 +\varepsilon}}\bigg) \partial_{S_B}\bigg(\frac{1}{r_2^{1 +\varepsilon}}\bigg)  + 1\leftrightarrow 2 \right)+  \propto  \frac{n_2^{S_A S_B}}{r_2^{2 +s_A+s_B+ 2\varepsilon}} \Bigg\} \, . 
\end{equation}
The first term in~\eqref{eq:SdVdV} is the source of the first term in \cref{eq:PdVdVsing}. It is then sufficient to show that applying on the former the Poisson integral in $d$ dimensions properly regularized at infinity, denoted as $\widetilde{\Delta}^{-1}$, cannot create a pole. Indeed, in the absence of pole, the limit of the result when $\varepsilon\to 0$ is a well-defined particular solution, which can only differ from the Hadamard's one by a regular homogeneous solution. This difference is also deprived of pole and reduces to zero if the same regularization is used to cure infra-red (IR) divergences at infinity in $d$ and three dimensions. To produce a pole in the current context, the $d$-dimensional Poisson integral operator $\Delta^{-1}$ has to be applied on contributions of the form $\hat{n}_1^{L}r_1^{\ell-2+q \varepsilon}$ or $\hat{n}_1^{L}r_1^{-\ell-3+q \varepsilon}$ with $q\in \mathbb{Z}$. On the other hand, since $s_{A,B}\geqslant 1$, our source term is made of elementary STF pieces $\hat{n}_1^L r_1^{\alpha +q\varepsilon}$ with $\alpha\leqslant -4$, which shows incidentally that its Poisson integral is well-defined at infinity, hence it is legitimate to work with the operator $\Delta^{-1}$ instead of $\widetilde{\Delta}^{-1}$. For this source term to contain a pole, we must have either $\alpha=\ell-2\leqslant -4$, which is impossible, or $\alpha=-\ell-3$ with $\ell \geqslant 1$. In that second situation, the parity $\pi_\sigma$ of the sum of the $\varepsilon^0$ coefficient in the power of $r_1$, namely $-\ell-3$, and of the number of $n_1^i$ factors, $\ell$, is odd, while the parity of the same sum, which we could refer to as $\pi_\sigma$-type parity, computed for $n_1^{S_AS_B}/r_1^{2+s_A+s_B+2\varepsilon}$ is even. This contradicts the fact that those two parities must be equal, and that, because the number of $n^i$ factors in $n^{S_AS_B}$ minus that in $\hat{n}^L$ is necessarily an even integer, \emph{i.e.}, $\ell$ and $s_A+s_B$ have the same parity. Thus, no pole can appear due to the action of $\widetilde{\Delta}^{-1}$ and it is safe to take the limit $\varepsilon \rightarrow 0$. As for the first term of \eqref{eq:PdVdVreg}, regular near $r_1=0$, its expansion consists of elementary pieces $\hat{r}_1^L r_1^{2k}$, with $k\in \mathbb{N}$, so none of them has the required form to produce a pole either.

The second terms in \eqref{eq:PdVdVsing} and \eqref{eq:PdVdVreg} are expressed in terms of the local functions \eqref{eq:gsing}--\eqref{eq:greg}. The limit $\varepsilon \to 0$ is well defined for each of them and commutes with the operators $R_1$ and $S_1$. The precise expression of $g^{(d)}_\text{reg}$ is unimportant for the analysis. The poles manifesting themselves in the sum over $\ell$ and in the last term of~\eqref{eq:greg} are mere constants, which are canceled after the action of the derivatives $\partial_{1S_A}\partial_{2S_B}$, with here $s_A+s_B\geqslant 2$. In fact, their combination even admits a finite limit $\varepsilon \to 0$, namely $\ln (r_{12}/r_0)-1$. The key point is that, due to $g_{\text{reg}}$, there is a non-zero regular part in the solution.

Note that the source~\eqref{eq:SdVdV} produces distributional pieces, \textit{e.g.}, $n_1^{S_A} r_1^{-n_A-1-\varepsilon} \partial^{n_B} (1/r_1^{1+\varepsilon})$, but those are zero provided $-\varepsilon$ is chosen to have a sufficiently large real part, thus vanishing in dimensional regularization. In the case $\varepsilon=0$, they are consistently discarded following Hadamard's regularization~\cite{BFreg}.

\subsection{Equivalence for the multipole moments}\label{app:proofmultipoles}

The multipole moments are defined as volume integrals over certain regular kernels, typically $\hat{x}_{ij}$, multiplied by compact sources (\textit{e.g.} $\sigma$), non-compact potentials, or derivatives of non-compact potentials. Since the structure of the involved potentials is known in $d$ dimensions, we are now in the position to investigate possible differences between the dimensional and Hadamard regularization arising from the volume integration.

\subsubsection{Compact support}

When the elementary source has a compact support, it can always be rewritten, using the same manipulation as for the transformation of the $\sigma$'s, as a sum of derivatives of $(F_\text{reg} \partial^{n_P} P \partial^{n_Q} Q ...)\delta^{(3)}_1$ (or likewise with $\delta^{(3)}_2$), where $P$, $Q$, ..., are compact-support or $\partial V \partial V$ potentials. Then, the factor in front of any given Dirac distribution can be substituted with its value at point 1, \textit{i.e.}, $\partial^{n_P} P\rightarrow (\partial^{n_P} P)_1$, which is nothing but $(\partial^{n_P} R_1[P])_1$ both for Hadamard and dimensional regularizations, according to the previous discussion. Since, in addition, $R_1[P]$ is continuous as a function of $\varepsilon$ in the limit $\varepsilon\to 0$, we conclude that the two prescriptions yield the same result for the integration of compact-support terms, with the convention that the Hadamard finite part $(~)_{1}$ of a product of derivative of potentials $\partial^{n_P} P \partial^{n_Q} Q\cdots$ should be ``distributive'', \textit{i.e.}, defined as $(\partial^{n_P} P)_{1} (\partial^{n_Q} Q)_{1}\cdots$.

\subsubsection{Non-compact support}

Let us next turn to the non-compact support terms. To get the UV difference $\mathcal{D}I$ between the integral $I^{(d)}$ over a sphere $|\mathbf{x}|<\mathcal{R}$ of a source $F^{(\varepsilon)}$ in $d=3+\varepsilon$ dimensions, with formal Taylor expansion near $\mathbf{x}=\boldsymbol{y_1}$,
\begin{align} \label{eq:diffdimreg}
F^{(\varepsilon)}(\mathbf{x}) = \sum \mathop{f}_{1} {}^{(\varepsilon)}_{p,q}(\boldsymbol{n}_1) \, r_1^{p+q\varepsilon}\,,
\end{align}
and its Hadamard counterpart $\int_{|\mathbf{x}|<\mathcal{R}} \dd^3 \mathbf{x}\, F(\mathbf{x})$, as $\varepsilon$ goes to 0, we resort to the formula~\cite{BDEI05dr}:
\begin{align}
  \mathcal{D}I= \frac{1}{\varepsilon}\sum_q \bigg[\frac{1}{q+1}+\varepsilon \ln s_1\bigg] \int \dd \Omega_{2+\varepsilon} \mathop{f}_1{}^{(\varepsilon)}_{-3,q} (\boldsymbol{n}_1)\, ,
\end{align}
where the sum over $q$ is finite. Based on this relation, we shall show that $\mathcal{D}I =0$, taking advantage of the fact that the angular integral of an odd number of $n_1^i$ is zero and that $\int \dd \Omega_{2+\varepsilon} \,\hat{n}_1^L \hat{n}_1^{P} = 0$ unless $\ell=p$. 

A close look at Eqs.~\eqref{eq:ILdecomp}--\eqref{eq:JLdecomp} reveals that all the terms but one are $\partial V \partial V$-type potentials multiplied by some regular functions. The only left term\footnote{As we have seen in \cref{eq:LaplaceF}, source terms of the type $\hat{x}^L \Delta F$ can be recast into surface integrals at infinity and are thus irrelevant for the present discussion about UV divergences.} is proportional to $\hat{x}_L \hat{W}_{ij}\partial_{ij} V$ at Newtonian order and requires a separate treatment. 

Let us consider now the pure $\partial V \partial V$-type terms, whose source structure at the leading Newtonian order required here is given by Eq.~\eqref{eq:SdVdV}. The derivatives in the product $\partial_{S_A} r_1^{-1-\varepsilon}\partial_{S_B} r_2^{-1-\varepsilon}$ comprise an ordinary and a distributional part which does non trivially vanish. The distributional part is made of derivatives of Dirac delta functions and is to be treated in the same way as compact support terms, this case having already been discussed above.

Regarding the ordinary parts of $\partial V \partial V$ terms, since evidently $\mathcal{D}I=0$ for regular functions, we focus on the singular part of these terms, which is only able to generate a difference. The first term of \cref{eq:SdVdV}, namely $\sum n_1^{S_A}n_1^{S_B}/ r_1^{2+s_A+s_B+2\varepsilon}$, with $s_{A,B}\geqslant 1$, has a non-zero coefficient $\mathop{f}_1{}^{(\varepsilon)}_{-3,q}$ in~\eqref{eq:diffdimreg} whenever $s_A+s_B+2=3$. This coefficient is therefore proportional to $n_1^i$ and its angular integral is zero. The second singular piece, of the form $\sum \partial_{S_A} r_1^{-1-\varepsilon} F_\text{reg}$, is made of a sum of terms $\propto r_1^{2k+\ell}\hat{n}_1^L\times r_1^{-1-s_A-\varepsilon} \hat{n}_1^{S_A}$ with $k\geqslant 0$, by virtue of Eq.~\eqref{eq:Freg}. The angular integral of $\mathop{f}_1{}^{(\varepsilon)}_{p,q}\propto \hat{n}_1^L\hat{n}_1^{S_A}$, for $p=-3=2k+\ell-1-s_A$, vanishes unless $\ell = s_A$, but this is impossible or, else, the contradictory statements $2k-1=-3$ and $k\geqslant 0$ would hold simultaneously. As a consequence, there is no contribution of the $\partial V\partial V$ sources to $\mathcal{D}I$.

Let us end with the ordinary part of the elementary integrand $\hat{x}_L \hat{W}_{ij}\partial_{ij} V$, with $\hat{W}_{ij}=\hat{W}^{(C)}_{ij}+\hat{W}^{(\partial V\partial V)}_{ij}$, truncated at Newtonian order. The contributions associated with the compact part $\hat{W}^{(C)}_{ij}$ or the distributional part have already been handled, so there only remains $\hat{x}_L \hat{W}^{(\partial V\partial V)}_{ij}\partial_{ij} V$, taken in the sense of functions. This product may be expanded into four pieces: 
\begin{itemize}
	\item[(i)] $\hat{x}_L \partial_{ij}V^{(C)}_\text{reg} (\hat{W}^{(\partial V\partial V)}_{ij})_\text{reg}$, which is purely regular [see \cref{eq:Freg}];
	\item[(ii)] $\hat{x}_L \partial_{ij}V^{(C)}_\text{sing} (\hat{W}^{(\partial V\partial V)}_{ij})_\text{reg}$, of the form $F_\text{reg}$ times \cref{eq:PCsing} (for $k=0$);
	\item[(iii)] $\hat{x}_L \partial_{ij}V^{(C)}_\text{reg} (\hat{W}^{(\partial V\partial V)}_{ij})_\text{sing}$, of the form $F_\text{reg}$ times \cref{eq:PdVdVsing};
	\item[(iv)] $\hat{x}_L (\hat{W}^{(\partial V\partial V)}_{ij})_\text{sing}\partial_{ij}V^{(C)}_\text{sing}$, of the form $F_\text{reg}$ times \cref{eq:PCsing} times \cref{eq:PdVdVsing}.
\end{itemize}
The piece (i) cannot contribute to $\mathcal{D}I$, while (ii) is structurally equal to $\partial_J r_1^{-1-\varepsilon}F_{\text{reg}}$, which has been already proved not to contribute to $\mathcal{D}I$ either. The piece (iii) can be decomposed into two parts, corresponding to the two terms under the curly brackets in Eq.~\eqref{eq:PdVdVsing}: 
\begin{itemize}
\item[(iiia)] The first part has a general term $\propto r_1^{2k+\ell}\hat{n}_1^L\times n_1^{J} r_1^{-s_A-s_B-2\varepsilon}$, for which the same parity argument as used in the $\partial V\partial V$ case applies: the $\pi_\sigma$-type parity is even, so that the angular integral of the $r_1^{-3}$ coefficient is necessarily zero.
\item[(iiib)] For the second part, the specific source of $V$ starts to play a role in the analysis. The number of derivatives $n$ in the leading tidal term of $\sigma$ is equal to two (both of space type), hence $V$ is schematically given by $\partial_{ab}(f_1(t)\delta_1) + 1\leftrightarrow 2$. This entails, for $\hat{W}_{ij}$, that $s_{A,B}= p_{A,B}+n_{A,B}\leqslant 3$ in the expansion~\eqref{eq:SdVdV} near $r_1=0$  in that case (with $p_A=p_B=1$), by virtue of Eq.~(4.2c). So, we find that $s_A\leqslant 3$ in the general term of (iii), namely $r_1^{2k+\ell}\hat{n}_1^L\times \partial_{1S_A} (r_{12}^{-1-\ell'-\varepsilon} r_1^{1+\ell'-\varepsilon}\hat{n}_1^{L'})$, which, therefore, cannot diverge faster than $r_1^{-2}$. Again, no contribution to $\mathcal{D}I$ arise.
\end{itemize}
The last piece, of type (iv), also consists of two sorts of terms:
\begin{itemize}
	\item[(iva)] The first ones, $\hat{x}^{ij}\times \partial_K r_1^{-1-\varepsilon}\times \hat{n}_1^Jr_1^{-s_A-s_B-2\varepsilon}$, turn out to be too divergent to contain $1/r_1^3$ powers in the tidal part. This is because we have for them either $k=4$ and $s_A+s_B\geqslant 2$, or $k\geqslant 2$ and $s_A+s_B= 4$, according to whether the tides originate from $V$ or $\hat{W}_{ij}$. In the point-particle part, only $\hat{r}_1^{ij} r_1^{-5-3\varepsilon}$ does not diverge faster than $r_1^{-3+q\varepsilon}$, but its angular integral vanishes. In both events, the corresponding $\mathcal{D}I$ is zero.
	\item[(ivb)] As for the second sort of terms, $F_\text{reg}\partial_I r_1^{-1-\varepsilon}\times \partial_{1S_A}\partial_{2S_B}g_{\text{sing}}^{(3+\varepsilon)}$, their $\pi_\sigma$-type parity is well defined and reduces to that of the general term of $r_1^{-1} g_{\text{sing}}^{(3)}$. It is thus even, preventing those terms to contribute to Eq.~\eqref{eq:diffdimreg}.
\end{itemize}

To conclude, we have shown that the Hadamard regularization is equivalent to the dimensional regularization when integrating the C source terms of the 2PN multipole moments that only involve compact-support and $\partial V\partial V$-type potentials. Likewise for the integration of NC sources of $\partial V\partial V$ type at leading order. We can also use Hadamard's regularization for the remaining non-compact support source $\propto \hat{W}_{ij}\partial_{ij}V$ at Newtonian order, provided the corrections beyond the point-particle model in $V$ are at least dipolar, which is indeed the case in our model [see the conditions on $k$ and $s_A+s_B$, in the above analysis of piece (iv)]. As a result, the Hadamard regularization is sufficient for all the computations presented in this paper.

\section{Tidal matter variables and CoM multipole moments}\label{app:tidal}

In \cref{sec:ready}, we have split the matter currents defined by~\eqref{eq:sigma} as $\sigma=\sigma_\text{pp}+\sigma_\text{tidal}$ \textit{etc.}, the point-particles parts being given by~\eqref{eq:pppart} together with the metric computed in \cref{sec:potentials}. The tidal parts are expressed in terms of the tidal moments $\hatG_{1ab}$, $\hatH_{1ab}$ and $\hatG_{1abc}$ projected onto the tetrad~\eqref{eq:tetrad} and evaluated at point 1 using the regularization, as 
\begin{subequations}\label{eq:sigmatidal}
	\begin{align}
	\sigma_\text{tidal}=&{}- \frac{1}{\sqrt{-g}} \partial_{ab}\bigg\{\delta_{1}{} \biggl(\mu_1^{(2)} \biggl [- \frac{1}{2} \hatG_{1}{}_{ab} + \frac{1}{c^2} (- \frac{3}{4} \hatG_{1}{}_{ab} v_{1}^{2} + \frac{3}{2} \hatG_{1}{}_{a}{}_{i} v_{1}^{b} v_{1}^{i} + \frac{1}{2} \hatG_{1}{}_{ab} V) \nonumber\\ & \qquad + \frac{1}{c^4} \Bigl(- \frac{7}{16} \hatG_{1}{}_{ab} v_{1}^{4} -  \frac{1}{8} (\hatG_{1}{}_{ij} v_{1}^{i} v_{1}^{j}) v_{1}^{a} v_{1}^{b} + \frac{7}{8} \hatG_{1}{}_{a}{}_{i} v_{1}^{2} v_{1}^{b} v_{1}^{i} -  \frac{1}{4} \hatG_{1}{}_{ab} v_{1}^{2} V + \frac{1}{2} \hatG_{1}{}_{a}{}_{i} V v_{1}^{b} v_{1}^{i} -  \frac{1}{4} \hatG_{1}{}_{ab} V^2 \nonumber\\ & \qquad \qquad + 2 \hatG_{1}{}_{ab} (v_{1}^{i} V_{i}) -  2\hatG_{1}{}_{a}{}_{i} v_{1}^{i} V_{b} -  2\hatG_{1}{}_{a}{}_{i} v_{1}^{b} V_{i} + \hatG_{1}{}_{bi} \hat{W}_{a}{}_{i} + \hatG_{1}{}_{ai} \hat{W}_{b}{}_{i}\Bigr)\biggl] \nonumber\\ & \quad + \sigma_1^{(2)} \Bigl(- \frac{4 \varepsilon_{a}{}_{ij} \hatH_{1}{}_{bj} v_{1}^{i}}{3c^2}  + \frac{1}{c^4} \Big(- \frac{2}{3} \varepsilon_{a}{}_{ij} \hatH_{1}{}_{bj} v_{1}^{2} v_{1}^{i}  + \frac{2}{3} \varepsilon_{a}{}_{jk} \hatH_{1}{}_{i}{}_{k} v_{1}^{b} v_{1}^{i} v_{1}^{j} + \frac{4}{3} \varepsilon_{a}{}_{ij} \hatH_{1}{}_{bj} V v_{1}^{i} + \frac{8}{3} \varepsilon_{a}{}_{ij} \hatH_{1}{}_{bj} V_{i} \Bigr)\biggl)\bigg\}\nonumber\\
	& -  \frac{1}{\sqrt{-g}} \biggl(\partial_t \partial_{a}\bigg\{\mu_1^{(2)} \delta_{1}{} \biggl [\frac{\hatG_{1}{}_{a}{}_{b} v_{1}^{b}}{c^2} + \frac{1}{c^4} \Bigl(\frac{1}{2} (\hatG_{1}{}_{ij} v_{1}^{i} v_{1}^{j}) v_{1}^{a} -  \hatG_{1}{}_{a}{}_{b} V v_{1}^{b}\Bigr)\biggl]\bigg\}\nonumber\\
	& + \partial_t \bigg\{\frac{\mu_1^{(2)} \delta_{1}{}}{c^4} \Bigl((\hatG_{1}{}_{ab} v_{1}^{a} \partial_{b}V) + 2 (\hatG_{1}{}_{ab} \partial_{b}V_{a})\Bigr)\bigg\}\biggl)\nonumber\\
	& -  \frac{1}{\sqrt{-g}} \partial_{a}\bigg\{\delta_{1}{} \biggl(\mu_1^{(2)} \biggl [- \frac{\hatG_{1}{}_{a}{}_{b} \partial_{b}V}{c^2} + \frac{1}{c^4} \Bigl(\hatG_{1}{}_{a}{}_{b} v_{1}^{b} \partial_t V + \frac{7}{2} (\hatG_{1}{}_{ij} v_{1}^{i} \partial_{j}V) v_{1}^{a} + \frac{7}{2} \hatG_{1}{}_{a}{}_{b} (v_{1}^{i} \partial_{i}V) v_{1}^{b} \nonumber\\ & \qquad \qquad - 2 (\hatG_{1}{}_{ij} v_{1}^{i} v_{1}^{j}) \partial_{a}V -  \frac{7}{2} \hatG_{1}{}_{a}{}_{b} v_{1}^{2} \partial_{b}V - 4 \hatG_{1}{}_{a}{}_{b} \partial_t V_{b} + 5 \hatG_{1}{}_{a}{}_{b} V \partial_{b}V + 4 (\hatG_{1}{}_{ij} \partial_{j}V_{i}) v_{1}^{a} + 2 \hatG_{1}{}_{bi} v_{1}^{b} \partial_{a}V_{i} \nonumber\\ & \qquad \qquad - 8 \hatG_{1}{}_{a}{}_{i} v_{1}^{b} \partial_{b}V_{i} - 2 \hatG_{1}{}_{bi} v_{1}^{b} \partial_{i}V_{a} +4 \hatG_{1}{}_{a}{}_{i} v_{1}^{b} \partial_{i}V_{b} + \hatG_{1}{}_{bi} \partial_{a}\hat{W}_{bi} - 2 \hatG_{1}{}_{bi} \partial_{i}\hat{W}_{a}{}_{b}\Bigr)\biggl] \nonumber\\ &\quad + \frac{\sigma_1^{(2)}}{c^4} \Big(\frac{8}{3} \varepsilon_{bij} \hatH_{1}{}_{aj} v_{1}^{b} \partial_{i}V -  \frac{8}{3} \varepsilon_{a}{}_{bj} \hatH_{1}{}_{i}{}_{j} v_{1}^{b} \partial_{i}V -  \frac{8}{3} \varepsilon_{bij} \hatH_{1}{}_{aj} \partial_{i}V_{b} -  \frac{8}{3} \varepsilon_{a}{}_{ij} \hatH_{1}{}_{b}{}_{j} \partial_{i}V_{b}\Big)\biggl)\bigg\}\nonumber\\
	& + \delta_{1}{} \biggl(\mu_1^{(2)} \biggl [\frac{1}{c^2} \Bigl(- (\hatG_{1}{}_{ab} \partial_{ab}V)
	+ \frac{3}{4} (\hatG_{1}{}_{ab} \hatG_{1}{}_{ab})\Bigr)
	+ \frac{1}{c^4} \Bigl(2 (\hatG_{1}{}_{ab} v_{1}^{a} \partial_t \partial_{b}V)
	+ (\hatG_{1}{}_{a}{}_{i} v_{1}^{a} v_{1}^{b} \partial_{ib}V)\nonumber\\ 
	& \qquad \qquad -  \frac{3}{2} (\hatG_{1}{}_{ab} \partial_{ab}V) v_{1}^{2}
	- 4 (\hatG_{1}{}_{ab} \partial_t \partial_{b}V_{a})
	+ 6 (\hatG_{1}{}_{ab} \partial_{a}V \partial_{b}V)
	+ 7 (\hatG_{1}{}_{ab} \partial_{ab}V) V
	- 4 (\hatG_{1}{}_{bi} v_{1}^{a} \partial_{ia}V_{b})\nonumber\\
	& \qquad \qquad + 4 (\hatG_{1}{}_{bi} v_{1}^{a} \partial_{ib}V_{a})
	+ \frac{9}{8} (\hatG_{1}{}_{ab} \hatG_{1}{}_{ab}) v_{1}^{2}
	-  \frac{3}{4} (\hatG_{1}{}_{ab} \hatG_{1}{}_{ab}) V\Bigr)\biggl] \nonumber\\ &
	+ \frac{\sigma_1^{(2)}}{c^4} \Bigl(- \frac{8}{3} (\varepsilon_{aij} \hatH_{1}{}_{bi} v_{1}^{a} \partial_{jb}V) -  \frac{16}{3} (\varepsilon_{bij} \hatH_{1}{}_{ab} \partial_{ja}V_{i})
	+ \frac{1}{2} (\hatH_{1}{}_{ab} \hatH_{1}{}_{ab})\Bigr)\biggl)
	-  \frac{1}{\sqrt{-g}} \partial_t^{2} \bigg\{\frac{\mu_1^{(2)} \delta_{1}{} (\hatG_{1}{}_{ab} v_{1}^{a} v_{1}^{b})}{2 c^4} \bigg\} \nonumber\\ & -  \frac{1}{\sqrt{-g}} \partial_{abi}\bigg\{\frac{1}{6} \mu_1^{(3)} \delta_{1}{} \hatG_{1}{}_{abi}\bigg\} + 1\leftrightarrow 2\, , \\[1ex]
	(\sigma_{i})_\text{tidal}={}&- \frac{1}{\sqrt{-g}} \biggl(\partial_t \partial_{a}\bigg\{\delta_{1}{} \biggl [\mu_1^{(2)} \Bigl(\frac{1}{2} \hatG_{1}{}_{ai} + \frac{1}{c^2} (\frac{1}{4} \hatG_{1}{}_{ai} v_{1}^{2} -  \frac{1}{4} \hatG_{1}{}_{i}{}_{b} v_{1}^{a} v_{1}^{b} -  \frac{1}{4} \hatG_{1}{}_{a}{}_{b} v_{1}^{b} v_{1}^{i} -  \frac{1}{2} \hatG_{1}{}_{ai} V)\Bigr) \nonumber\\ & \quad + \frac{\sigma_1^{(2)}}{c^2} \Big(\frac{1}{3} \varepsilon_{i}{}_{bj} \hatH_{1}{}_{aj} v_{1}^{b} + \frac{1}{3} \varepsilon_{a}{}_{bj} \hatH_{1}{}_{ij} v_{1}^{b}\Big)\biggl]\bigg\} + \partial_t \bigg\{\frac{\mu_1^{(2)} \delta_{1}{} \hatG_{1}{}_{i}{}_{a} \partial_{a}V}{c^2} \bigg\}\biggl)\nonumber\\
	& -  \frac{1}{\sqrt{-g}} \partial_{ab}\bigg\{\delta_{1}{} \biggl [\mu_1^{(2)} \Bigl(\frac{1}{2} \hatG_{1}{}_{ai} v_{1}^{b}-  \frac{1}{2} \hatG_{1}{}_{ab} v_{1}^{i} + \frac{1}{c^2} \Big(\frac{1}{4} \hatG_{1}{}_{ai} v_{1}^{2} v_{1}^{b}-  \frac{1}{4} \hatG_{1}{}_{ab} v_{1}^{2} v_{1}^{i} -  \frac{1}{4} \hatG_{1}{}_{i}{}_{j} v_{1}^{a} v_{1}^{b} v_{1}^{j} + \frac{1}{4} \hatG_{1}{}_{a}{}_{j} v_{1}^{b} v_{1}^{i} v_{1}^{j} \nonumber\\ & \qquad \qquad -  \frac{1}{2} \hatG_{1}{}_{ai} V v_{1}^{b} + \frac{1}{2} \hatG_{1}{}_{ab} V v_{1}^{i}\Big)\Bigr) \nonumber\\ &+ \sigma_1^{(2)} \Bigl(- \frac{1}{3} \varepsilon_{ai}{}_{j} \hatH_{1}{}_{bj} + \frac{1}{c^2} \Big(- \frac{1}{6} \varepsilon_{ai}{}_{j} \hatH_{1}{}_{bj} v_{1}^{2} + \frac{1}{3} \varepsilon_{a}{}_{jk} \hatH_{1}{}_{ik} v_{1}^{b} v_{1}^{j} + \frac{1}{6} \varepsilon_{ai}{}_{k} \hatH_{1}{}_{j}{}_{k} v_{1}^{b} v_{1}^{j} -  \frac{1}{3} \varepsilon_{a}{}_{jk} \hatH_{1}{}_{bk} v_{1}^{i} v_{1}^{j} \nonumber\\ &\qquad \qquad+ \frac{1}{6} \varepsilon_{ai}{}_{k} \hatH_{1}{}_{b}{}_{j} v_{1}^{j} v_{1}^{k} +  \varepsilon_{ai}{}_{j} \hatH_{1}{}_{bj} V \Big)\Bigr)\biggl]\bigg\}\nonumber\\
	& + \delta_{1}{} \biggl [\frac{2 \sigma_1^{(2)} \varepsilon_{i}{}_{bj} \hatH_{1}{}_{ab} \partial_{ja}V}{3 c^2}
	+ \frac{\mu_1^{(2)}}{c^2} \Bigl(- \frac{1}{2} \hatG_{1}{}_{ib} v_{1}^{a} \partial_{ab}V
	-  \frac{1}{2} \hatG_{1}{}_{a}{}_{b} v_{1}^{a} \partial_{ib}V
	+ \frac{3}{4} (\hatG_{1}{}_{ab} \hatG_{1}{}_{ab}) v_{1}^{i}\Bigr)\biggl]\nonumber\\
	& -  \frac{1}{\sqrt{-g}} \partial_{a}\bigg\{\delta_{1}{} \biggl [\frac{\sigma_1^{(2)}}{c^2} \Big(- \frac{2}{3} \varepsilon_{ai}{}_{j} \hatH_{1}{}_{b}{}_{j} \partial_{b}V -  \frac{2}{3} \varepsilon_{a}{}_{bj} \hatH_{1}{}_{ij} \partial_{b}V\Big) \nonumber\\ &\quad + \frac{\mu_1^{(2)}}{c^2} \Bigl(- \frac{1}{2} \hatG_{1}{}_{ai} \partial_t V -  \hatG_{1}{}_{ai} (v_{1}^{j} \partial_{j}V) -  \frac{1}{2} \hatG_{1}{}_{i}{}_{b} v_{1}^{b} \partial_{a}V + 2 \hatG_{1}{}_{i}{}_{b} v_{1}^{a} \partial_{b}V -  \hatG_{1}{}_{a}{}_{b} v_{1}^{i} \partial_{b}V + \frac{1}{2} \hatG_{1}{}_{a}{}_{b} v_{1}^{b} \partial_{i}V \nonumber\\ &\qquad \qquad + \hatG_{1}{}_{i}{}_{b} \partial_{a}V_{b} -  \hatG_{1}{}_{i}{}_{b} \partial_{b}V_{a} + \hatG_{1}{}_{a}{}_{b} \partial_{b}V_{i} -  \hatG_{1}{}_{a}{}_{b} \partial_{i}V_{b}\Bigr)\biggl]\bigg\} + 1\leftrightarrow 2\, , \label{eq:sigmatidali}\\[1ex]
	(\sigma_{ij})_\text{tidal}={}&- \frac{1}{\sqrt{-g}} \partial_t^{2} \bigg\{- \frac{1}{2} \mu_1^{(2)} \delta_{1}{} \hatG_{1}{}_{ij}\bigg\} - \mu_1^{(2)} \delta_{1}{} \hatG_{1}{}_{a(i} \partial_{j)a}V  -  \frac{1}{\sqrt{-g}} \partial_{a}\bigg\{\mu_1^{(2)} \delta_{1}{} \Big(\frac{1}{2} \hatG_{1}{}_{ij} \partial_{a}V -  \hatG_{1}{}_{a(i} \partial_{j)}V \Big)\bigg\}\nonumber\\
	& -  \frac{1}{\sqrt{-g}} \partial_t \partial_{a}\bigg\{\delta_{1}{} \Bigl(\mu_1^{(2)} (- \hatG_{1}{}_{ij} v_{1}^{a} + \hatG_{1}{}_{a(i} v_{1}^{j)} ) - \frac{2\sigma_1^{(2)}  \varepsilon_{ab(i} \hatH_{1}{}_{j)b}}{3}\Bigr)\bigg\}\nonumber\\
	& -  \frac{1}{\sqrt{-g}} \partial_{ab}\bigg\{\delta_{1}{} \Bigl(\mu_1^{(2)} \Big(- \frac{1}{2} \hatG_{1}{}_{ij} v_{1}^{a} v_{1}^{b} + \hatG_{1}{}_{a(i} v_{1}^{j)}v_{1}^{b}-  \frac{1}{2} \hatG_{1}{}_{ab} v_{1}^{i} v_{1}^{j}\Big) \nonumber\\ & \quad + \sigma_1^{(2)} \Big(\frac{2}{3} \varepsilon_{ka(i} \hatH_{1}{}_{j)k} v_{1}^{b}  -  \frac{2}{3} \varepsilon_{ka(i} v_{1}^{j)}\hatH_{1}{}_{bk} \Big)\Bigr)\bigg\} + 1\leftrightarrow 2\,.
	\end{align}
\end{subequations}

In \cref{sec:MultipolarMoments}, the multipole moments have been computed to NNL order. We provide here their tidal parts in the frame of the CoM: 
\begin{subequations}\label{eq:momentsCM}
	\begin{align}
	I^{\text{tidal}}_{ij}={}&\frac{G m}{r^3} \Bigg\{ n^{\langle i} n^{j\rangle} \Bigl(3 \mu_{+}^{(2)}
	+ 3 \Delta\mu_{-}^{(2)}\Bigr)
	+ \frac{1}{c^{2}} \Biggl[n^{\langle i} n^{j\rangle} \biggl\{\biggl [\mu_{+}^{(2)} \Bigl(- \frac{15}{2}
	+ \frac{250}{7} \nu \Bigr)
	+ \Delta\mu_{-}^{(2)} \Bigl(- \frac{15}{2}
	-  \frac{40}{7} \nu \Bigr) + \frac{160}{3} \nu \sigma_{+}^{(2)}\biggl] \dot{r}^2\nonumber\\
	& \qquad  
	+ \biggl [\mu_{+}^{(2)} \Bigl(6
	+ \frac{13}{7} \nu \Bigr)
	+ \Delta\mu_{-}^{(2)} \Bigl(6
	-  \frac{13}{7} \nu \Bigr)
	+ \frac{160}{3} \nu \sigma_{+}^{(2)}\biggl] v^{2}
	+ \frac{G m}{r} \biggl [\mu_{+}^{(2)} \Bigl(- \frac{15}{2}
	-  \frac{99}{7} \nu + \frac{39}{7} \nu^2\Bigr)\nonumber\\
	& \qquad \quad 
	+ \Delta\mu_{-}^{(2)} \Bigl(- \frac{15}{2}
	-  \frac{54}{7} \nu \Bigr)\biggl]\biggl\}
	+ n^{\langle i} v^{j\rangle} \biggl [\mu_{+}^{(2)} \Bigl(-6
	-  \frac{214}{7} \nu \Bigr)
	+ \Delta\mu_{-}^{(2)} \Bigl(-6
	+ \frac{130}{7} \nu \Bigr)
	-  \frac{448}{3} \nu \sigma_{+}^{(2)}\biggl] \dot{r}\nonumber\\
	& \quad + v^{\langle i} v^{j\rangle} \biggl [\mu_{+}^{(2)} \Bigl(3
	+ \frac{38}{7} \nu \Bigr)
	+ \Delta\mu_{-}^{(2)} \Bigl(3
	-  \frac{38}{7} \nu \Bigr)
	+ \frac{128}{3} \nu \sigma_{+}^{(2)}\biggl]\Biggl]
	\nonumber\\
	&+ \frac{1}{c^4} \Bigg\{ n^{\langle i} n^{j\rangle} \Biggl[\biggl [\mu_{+}^{(2)} \Bigl(\frac{105}{8}
	- 20 \nu - 85 \nu^2\Bigr)
	+ \Delta\mu_{-}^{(2)} \Bigl(\frac{105}{8}
	- 85 \nu
	- 15 \nu^2\Bigr)
	+ \sigma_{+}^{(2)} \Bigl(\frac{80}{3} \nu
	-  \frac{160}{3} \nu^2\Bigr)
	-  \frac{640}{3} \Delta\sigma_{-}^{(2)} \nu \biggl] \dot{r}^4
	\nonumber\\
	&\qquad + \biggl [\mu_{+}^{(2)} \Bigl(- \frac{45}{2}
	+ 120 \nu -  \frac{515}{14} \nu^2\Bigr)
	+ \Delta\mu_{-}^{(2)} \Bigl(- \frac{45}{2}
	+ 15 \nu
	+ \frac{205}{14} \nu^2\Bigr)
	+ \sigma_{+}^{(2)} \Bigl(\frac{600}{7} \nu
	-  \frac{640}{7} \nu^2\Bigr)
	-  \frac{600}{7} \Delta\sigma_{-}^{(2)} \nu \biggl] \dot{r}^2 v^{2}
	\nonumber\\
	& \qquad + \biggl [\mu_{+}^{(2)} \Bigl(6-  \frac{72}{7} \nu
	-  \frac{47}{14} \nu^2\Bigr)
	+ \Delta\mu_{-}^{(2)} \Bigl(6
	-  \frac{96}{7} \nu
	+ \frac{61}{14} \nu^2\Bigr)
	+ \sigma_{+}^{(2)} \Bigl(\frac{424}{7} \nu
	-  \frac{736}{7} \nu^2\Bigr)
	+ \frac{136}{7} \Delta\sigma_{-}^{(2)} \nu \biggl] v^{4}\nonumber\\
	& \qquad + \frac{G m}{r} \biggl\{\biggl [\mu_{+}^{(2)} \Bigl(- \frac{639}{8}
	+ \frac{22885}{56} \nu
	+ \frac{3539}{56} \nu^2
	-  \frac{213}{7} \nu^3\Bigr)
	+ \Delta\mu_{-}^{(2)} \Bigl(- \frac{639}{8}
	+ \frac{9565}{56} \nu
	-  \frac{4909}{56} \nu^2\Bigr)\nonumber\\
	& \qquad \qquad + \sigma_{+}^{(2)} \Bigl(- \frac{164}{7} \nu
	+ \frac{2600}{21} \nu^2\Bigr)
	-  \frac{340}{21} \Delta\sigma_{-}^{(2)} \nu \biggl] \dot{r}^2
	+ \biggl [\mu_{+}^{(2)} \Bigl(\frac{139}{8}
	-  \frac{14405}{168} \nu
	-  \frac{1537}{168} \nu^2
	-  \frac{953}{42} \nu^3\Bigr)\nonumber\\
	& \qquad \qquad + \Delta\mu_{-}^{(2)} \Bigl(\frac{139}{8}
	-  \frac{3211}{56} \nu
	+ \frac{709}{24} \nu^2\Bigr)
	+ \sigma_{+}^{(2)} \Bigl(\frac{4804}{63} \nu
	-  \frac{2248}{21} \nu^2\Bigr)
	+ \frac{508}{63} \Delta\sigma_{-}^{(2)} \nu \biggl] v^{2}\biggl\}
	\nonumber\\ 
	& \qquad + \frac{G^2 m^2}{r^2} \biggl [\mu_{+}^{(2)} \Bigl(\frac{285}{28}+ \frac{8885}{84} \nu
	-  \frac{3147}{28} \nu^2
	+ \frac{27}{2} \nu^3\Bigr)
	+ \Delta\mu_{-}^{(2)} \Bigl(\frac{285}{28}
	+ \frac{8047}{84} \nu
	+ \frac{233}{84} \nu^2\Bigr)\biggl]\Biggl]
	\nonumber\\
	& \quad + n^{\langle i} v^{j\rangle} \biggl\{\biggl [\mu_{+}^{(2)} \Bigl(15
	-  \frac{1265}{7} \nu+ \frac{1810}{7} \nu^2\Bigr)
	+ \Delta\mu_{-}^{(2)} \Bigl(15
	+ \frac{1055}{7} \nu
	+ \frac{110}{7} \nu^2\Bigr)
	\nonumber \\ &\qquad \qquad + \sigma_{+}^{(2)} \Bigl(- \frac{560}{3} \nu
	+ \frac{640}{3} \nu^2\Bigr) 
	+ \frac{1840}{3} \Delta\sigma_{-}^{(2)} \nu \biggl] \dot{r}^3
	+ \biggl [\mu_{+}^{(2)} \Bigl(-6
	+ \frac{15}{7} \nu
	-  \frac{346}{7} \nu^2\Bigr)\nonumber\\ & \qquad \qquad 
	+ \Delta\mu_{-}^{(2)} \Bigl(-6
	+ \frac{27}{7} \nu
	-  \frac{286}{7} \nu^2\Bigr)
	+ \sigma_{+}^{(2)} \Bigl(- \frac{3392}{21} \nu
	+ \frac{5440}{21} \nu^2\Bigr)
	-  \frac{2432}{21} \Delta\sigma_{-}^{(2)} \nu \biggl] \dot{r} v^{2}\nonumber\\
	& \qquad + \frac{G m}{r} \biggl [\mu_{+}^{(2)} \Bigl(\frac{105}{2}
	-  \frac{10601}{42} \nu
	-  \frac{569}{14} \nu^2
	+ \frac{16}{7} \nu^3\Bigr)
	+ \Delta\mu_{-}^{(2)} \Bigl(\frac{105}{2}
	-  \frac{2329}{42} \nu
	-  \frac{731}{42} \nu^2\Bigr)
	\nonumber\\
	& \qquad \qquad + \sigma_{+}^{(2)} \Bigl(-96 \nu+ \frac{1312}{21} \nu^2\Bigr)
	+ \frac{256}{21} \Delta\sigma_{-}^{(2)} \nu \biggl] \dot{r}\biggl\}
	\nonumber\\ & \quad + v^{\langle i} v^{j\rangle} \biggl\{\biggl [\mu_{+}^{(2)} \Bigl(- \frac{15}{2}
	+ \frac{687}{7} \nu
	-  \frac{760}{7} \nu^2\Bigr)
	+ \Delta\mu_{-}^{(2)} \Bigl(- \frac{15}{2}
	-  \frac{477}{7} \nu
	+ \frac{20}{7} \nu^2\Bigr)\nonumber\\
	& \qquad \qquad + \sigma_{+}^{(2)} \Bigl(\frac{3104}{21} \nu
	-  \frac{3520}{21} \nu^2\Bigr)
	-  \frac{5344}{21} \Delta\sigma_{-}^{(2)} \nu \biggl] \dot{r}^2
	+ \biggl [\mu_{+}^{(2)} \Bigl(3
	- 11 \nu
	+ \frac{109}{7} \nu^2\Bigr)
	+ \Delta\mu_{-}^{(2)} \Bigl(3
	-  \nu
	+ \frac{65}{7} \nu^2\Bigr)\nonumber\\
	& \qquad \qquad + \sigma_{+}^{(2)} \Bigl(\frac{192}{7} \nu
	-  \frac{384}{7} \nu^2\Bigr)
	+ \frac{256}{7} \Delta\sigma_{-}^{(2)} \nu \biggl] v^{2}
	\nonumber\\ & \qquad + \frac{G m}{r} \biggl [\mu_{+}^{(2)} \Bigl(- \frac{7}{2}
	+ \frac{458}{21} \nu
	+ \frac{101}{21} \nu^2
	-  \frac{485}{21} \nu^3\Bigr)
	+ \Delta\mu_{-}^{(2)} \Bigl(- \frac{7}{2}
	-  \frac{151}{21} \nu
	+ \frac{332}{21} \nu^2\Bigr)
	\nonumber\\ & \qquad \qquad+ \sigma_{+}^{(2)} \Bigl(\frac{2720}{63} \nu
	-  \frac{1664}{21} \nu^2\Bigr)
	-  \frac{256}{63} \Delta\sigma_{-}^{(2)} \nu \biggl]\biggl\}\Bigg\}\Bigg\}\,,\\
	I^{\text{tidal}}_{ijk}={}&\frac{G m \nu}{r^2} \Bigg\{ 18 \mu_{-}^{(2)} n^{\langle i} n^{j} n^{k\rangle}
	+ \frac{1}{c^{2}} \Biggl[n^{\langle i} n^{j} n^{k\rangle} \biggl\{\biggl [
	-  \frac{105}{2} \Delta\mu_{+}^{(2)}
	+ \mu_{-}^{(2)} \Bigl(\frac{15}{2}
	- 15 \nu \Bigr) -60 \Delta\sigma_{+}^{(2)} +60 \sigma_{-}^{(2)}\biggl] \dot{r}^2\nonumber\\
	& 
	\qquad + \biggl [- 3 \Delta\mu_{+}^{(2)}
	+ \mu_{-}^{(2)} \Bigl(48
	- 33 \nu \Bigr) -84 \Delta\sigma_{+}^{(2)}
	+ 84 \sigma_{-}^{(2)}\biggl] v^{2}
	\nonumber\\
	& \qquad + \frac{G m}{r} \biggl [\Delta\mu_{+}^{(2)} \Bigl(\frac{33}{2}
	- 9 \nu \Bigr)+ \mu_{-}^{(2)} \Bigl(- \frac{135}{2}
	- 27 \nu \Bigr)\biggl]\biggl\}\nonumber\\ &
	\quad + n^{\langle i} n^{j} v^{k\rangle} \biggl [ 72 \Delta\mu_{+}^{(2)}
	+ \mu_{-}^{(2)} \Bigl(-108
	+ 108 \nu \Bigr)+216 \Delta\sigma_{+}^{(2)}
	- 216 \sigma_{-}^{(2)}\biggl] \dot{r}\nonumber\\
	& \quad + n^{\langle i} v^{j} v^{k\rangle} \biggl [
	- 21 \Delta\mu_{+}^{(2)}
	+ \mu_{-}^{(2)} \Bigl(39
	- 42 \nu \Bigr) -72 \Delta\sigma_{+}^{(2)}
	+ 72 \sigma_{-}^{(2)}\biggl]\Biggl]\Bigg\}\,,\\
	I^{\text{tidal}}_{ijkl}={}&\frac{G m \nu}{r} n^{\langle i} n^{j} n^{k} n^{l \rangle} \Bigl(18 \mu_{+}^{(2)}
	- 18 \Delta\mu_{-}^{(2)}\Bigr)\,,\\
	J^{\text{tidal}}_{ij}={}&\frac{G m}{r^3} \Bigg\{ n^{\langle i} (n\times v)^{j\rangle} \Bigl( 9 \nu \mu_{-}^{(2)} + 12 \Delta\sigma_{+}^{(2)}
	+ 12 \sigma_{-}^{(2)}\Bigr) \nonumber\\ &
	+ \frac{1}{c^{2}} \Biggl[n^{\langle i} (n\times v)^{j\rangle} \biggl\{\biggl [
	-  \frac{240}{7} \Delta\mu_{+}^{(2)} \nu
	+ \mu_{-}^{(2)} \Bigl(\frac{375}{14} \nu
	-  \frac{60}{7} \nu^2\Bigr) +\Delta\sigma_{+}^{(2)} \Bigl(-30
	-  \frac{360}{7} \nu \Bigr)
	+ \sigma_{-}^{(2)} \Bigl(-30
	+ \frac{1200}{7} \nu \Bigr)\biggl] \dot{r}^2
	\nonumber\\
	& \qquad + \biggl [
	 \frac{129}{28} \Delta\mu_{+}^{(2)} \nu + \mu_{-}^{(2)} \Bigl(\frac{417}{28} \nu
	-  \frac{51}{7} \nu^2\Bigr) + \Delta\sigma_{+}^{(2)} \Bigl(12
	-  \frac{96}{7} \nu \Bigr)
	+ \sigma_{-}^{(2)} \Bigl(12
	-  \frac{72}{7} \nu \Bigr)\biggl] v^{2}
	\nonumber\\ &\qquad + \frac{G m}{r} \biggl [
	\Delta\mu_{+}^{(2)} \Bigl(\frac{3}{14} \nu
	-  \frac{60}{7} \nu^2\Bigr) + \mu_{-}^{(2)} \Bigl(- \frac{111}{14} \nu
	-  \frac{267}{7} \nu^2\Bigr) + \Delta\sigma_{+}^{(2)} \Bigl(-4
	-  \frac{300}{7} \nu \Bigr)
	+ \sigma_{-}^{(2)} \Bigl(-4
	-  \frac{316}{7} \nu \Bigr)\biggl]\biggl\}\nonumber\\
	& 
	\quad + (n\times v)^{\langle i} v^{j\rangle} \biggl [
	\frac{213}{14} \Delta\mu_{+}^{(2)} \nu
	+ \mu_{-}^{(2)} \Bigl(- \frac{507}{14} \nu
	+ \frac{129}{7} \nu^2\Bigr) +\frac{396}{7} \Delta\sigma_{+}^{(2)} \nu
	-  \frac{564}{7} \nu \sigma_{-}^{(2)}\biggl] \dot{r}\Biggl]\Bigg\}\,,\\
	J^{\text{tidal}}_{ijk}={}&\frac{G m \nu}{r^2} n^{\langle j} n^{k} (n\times v)^{i\rangle} \Bigl(12 \mu_{+}^{(2)}
	- 12 \Delta\mu_{-}^{(2)}
	+ 64 \sigma_{+}^{(2)}\Bigr)\,.
	\end{align}
\end{subequations}

\bibliography{ListeRef_HFB20}

\end{document}